\documentclass[prx,twocolumn,notitlepage,superscriptaddress]{revtex4-2}
\usepackage{amsmath}
\usepackage{amssymb}
\usepackage{graphicx}
\usepackage{comment}
\usepackage[caption=false]{subfig}
\usepackage{wasysym}
\usepackage{cancel}

\renewcommand{\k}{\mathbf{k}}
\newcommand{\q}{\mathbf{q}}

\renewcommand{\r}{\mathbf{r}}
\newcommand{\R}{\mathbf{R}}

\newcommand{\ND}[1]{\hat{n}_{#1}}

\newcommand{\DD}[1]{\OPc{d}{#1}}
\newcommand{\D}[1]{\OP{d}{#1}}

\newcommand{\captiontitle}[1]{\textbf{#1}}

\newcommand{\Aberry}{\boldsymbol{\mathcal{A}}}

\newcommand{\Agauge}{\mathbf{A}}

\newcommand{\sectiontitle}[1]{\subsection*{#1}}

\newcommand{\Ham}{\hat{H}}

\newcommand{\ketbra}[2]{\left|#1\middle>\middle<#2\right|}
\newcommand{\braket}[2]{\left<#1\middle|#2\right>}

\newcommand{\bra}[1]{\left<#1\right|}
\newcommand{\ket}[1]{\left|#1\right>}
\newcommand{\OPc}[2]{\hat{#1}_{#2}^{\dag}}
\newcommand{\OP}[2]{\hat{#1}_{#2}^{\vphantom{\dag}}}
\newcommand{\CD}[1]{\OPc{c}{#1}}
\newcommand{\C}[1]{\OP{c}{#1}}

\newcommand{\hc}{\textrm{h.c.}}
\newcommand{\E}{\epsilon}

\DeclareMathOperator{\Real}{Re}

\begin{document}
\title{Quantum-Geometric Light-Matter Coupling in Correlated Quantum Materials}
\date{\today}
\author{Wai Ting Tai}
\affiliation{Department of Physics and Astronomy, University of Pennsylvania, Philadelphia, PA 19104}
\author{Martin Claassen}
\email{claassen@sas.upenn.edu}
\affiliation{Department of Physics and Astronomy, University of Pennsylvania, Philadelphia, PA 19104}

\begin{abstract}

Irradiation with light provides a powerful tool to interrogate, control or induce new quantum states of matter out of equilibrium, however a microscopic understanding of light-matter coupling in interacting electron systems remains a profound challenge. Here, we show that light grants a new quantum-geometric handle to steer and probe correlated quantum materials, whereby photons can couple directly to the shape and center of the maximally-localized Wannier functions that comprise the material's interacting bands, dressing both electronic motion and electronic interactions with light. Notably, this effect is generic to any material and purely geometric in origin, but dominates emergent optical responses in correlated electron systems with poorly-localized or obstructed Wannier functions. Spectroscopic consequences are first illustrated for a paradigmatic strongly-interacting model with a tunable Wannier obstruction. We then present ramifications for non-equilibrium control of moir\'e heterostructures and find that subjecting magic-angle twisted bilayer graphene to weak THz radiation can conspire with a fragile topological obstruction to profoundly alter the material's competing interactions and tune across boundaries to competing phases. 
\end{abstract}
\maketitle

\sectiontitle{Introduction}
Understanding how photons couple to quantum matter is foundational to interrogating and manipulating materials with light. With free-electron optical responses largely well understood, the past decade has seen rapid advances in photon-based and time-resolved spectroscopies to probe and control strongly-correlated electron systems \cite{delatorre21,sobota21,wang18} which are in turn found to host an ever expanding panoply of quantum phases with novel and useful properties. However, while these phases are well-captured in equilibrium to emerge from an intricate competition between electronic motion and interactions within only a \textit{small open-shell set} of the material's valence orbitals \cite{HubbardManyElectronCollaboration2015,FutureOfTheCorrelatedElectronProblem2020,DFTDMFT2019}, remarkably no such simple understanding exists for coupling these degrees of freedom to light.

Fundamentally, the photon frequency grants a natural handle to target excitations at select energy scales. Some of these -- such as optical interband transitions or core-hole excitations -- involve select additional degrees of freedom; conversely, one should expect on physical grounds that \textit{low-frequency} photons -- typically, in the terahertz regime -- should couple only to interacting electrons within the relevant valence states near the Fermi energy, while leaving filled deeper valence bands and higher-energy empty conduction bands inert. A microscopic understanding of this \textit{effective} coupling between light and correlated electrons is crucial to interpret spectroscopic responses, to understand what THz spectroscopy actually measures, and to steer correlated phases with strong driving fields into non-equilibrium states with new and useful properties. However, its study poses a surprising theoretical challenge for strongly-interacting quantum materials.

Indeed, formalizing the above intuition is immediately fraught with conceptual difficulties: While light-matter coupling in solids can be conventionally thought of as the field coupling to a current (in velocity gauge) \cite{MahanBook}, the effective electron velocity can be suppressed or even vanish in correlated flat band compounds \cite{regnault21} although the intraband coupling to THz photons should not. Similarly, strong fields are commonly described to couple to interacting electrons via a Peierls phase accrued upon hopping between neighboring valence orbitals in a crystal \cite{peierls33,kohn1959}, with recent works introducing and benchmarking corrections to account for dipole transitions in multi-orbital models for classical and quantized light \cite{golez19,golez19b,li20,dmytruk21}. However, a faithful description of the active valence bands of a solid naturally leads to Wannier orbitals with a finite spatial extent \cite{marzari11}, precluding a sharp definition of the hopping trajectory. Simultaneously, a finite Wannier spread is intimately tied to finite-range electron-electron scattering even for \textit{a priori} well-screened Coulomb interactions, which, even if irrelevant in equilibrium, must acquire a Peierls phase and couple to light as well. While the above obstacles apply to all quantum materials, they are exacerbated for partially-filled obstructed or topological bands, which do not admit a representation in terms of local Wannier functions at all \cite{brouder07,po2018}. This prominently includes moir\'e heterostructures of two-dimensional van der Waals materials such as twisted bilayer graphene (TBG) or twisted transition-metal dichalcogenides \cite{balents20,kennes21,andrei20,andrei21,kang2018,kang2019,koshino2018,po2019,zou18,carr2019,wang19, Devakul2021, Luo2022} -- systems with almost dispersionless bands composed of charge puddles trapped by the moir\'e superlattice potential which realize a myriad of interaction-driven phases.

Theoretically, apparent discrepancies between electromagnetic gauge choices provide a complementary perspective of the problem \cite{foreman02}. Quantum mechanics prescribes that electrons couple to the gauge-dependent vector ($\mathbf{A}$) and scalar ($\phi$) potentials as opposed to the physical electric and magnetic fields. However, while local gauge invariance is trivially satisfied by the bare many-body Hamiltonian, the equivalence between different gauge choices becomes exceedingly hard to retain in effective theories of the bands near the Fermi energy, with well-known repercussions arising already in calculations of free-electron optical responses in insulators, which vanish as expected at zero frequency in length gauge ($\mathbf{A}=0$) while exhibiting spurious low-frequency divergences in velocity gauge ($\phi=0$) calculations. The former have been successfully employed to calculate non-linear optical susceptibilities \cite{aversa1995,sipe2000}. However, the former do not readily extend to periodic many-body systems, whereas sum rules for the latter dictate that these divergences cancel if and only if all the solid's bands are included in the calculation \cite{ventura2017,taghizadeh2017,passos2018,parker2019} -- a task that is unfeasible for many-body systems and stands in stark contrast with the expectation that low-frequency responses should involve only low-energy degrees of freedom of a solid. Similar considerations apply to gauge-invariant Peierls-coupled multiband tight-binding models; for instance, quantum-metric contributions to light-matter coupling in TBG have recently been proposed to arise after projecting to flat bands near charge neutrality \cite{topp2021}, but can instead be shown to erroneously arise as a result of gauge-invariance-breaking terms that again lead to spurious low-frequency divergences. Conversely, para- and diamagnetic current operators for an individual topologically-trivial flat band were shown to involve solely interaction-mediated contributions \cite{mao22}. Similarly, careful \textit{ab initio} benchmarks have recently shed light on failures of light-matter coupled tight-binding descriptions constrained to the active bands \cite{taghizadeh2017,taghizadeh2018}, and possible \textit{ad hoc} corrections to cure low-frequency divergences have been proposed \cite{schuler2021}.

In this work, we present a new quantum-geometric interaction of light and matter in correlated materials, whereby photons directly deform the shape and motion of the Wannier orbitals that comprise a quantum material's bands near the Fermi energy, thereby dressing both electronic motion and electronic interactions with light. This mechanism is generic to all materials and purely geometric in origin, complements the usual Peierls phase in velocity gauge calculations, but dominates optical driving and responses for strongly-interacting electron systems with poorly localized Wannier orbitals. This includes materials with finite Berry curvature or a finite quantum geometric tensor, and prominently encompasses moir\'e heterostructures and interacting topological materials. We first illustrate spectroscopic consequences by example of a strongly-interacting topological one-dimensional chain, which reveals a quantum-geometric coupling of photons to an emergent domain wall mode which can be interrogated and melted with light. We then present ramifications for non-equilibrium control of moir\'e heterostructures by subjecting magic-angle twisted bilayer graphene to weak THz radiation, which conspires with a fragile topological obstruction to profoundly alter the material's competing interactions, thereby granting a new photon-based knob to tune TBG across boundaries to competing phases.

\begin{figure}[t]
	\centering
	\includegraphics[width=\columnwidth]{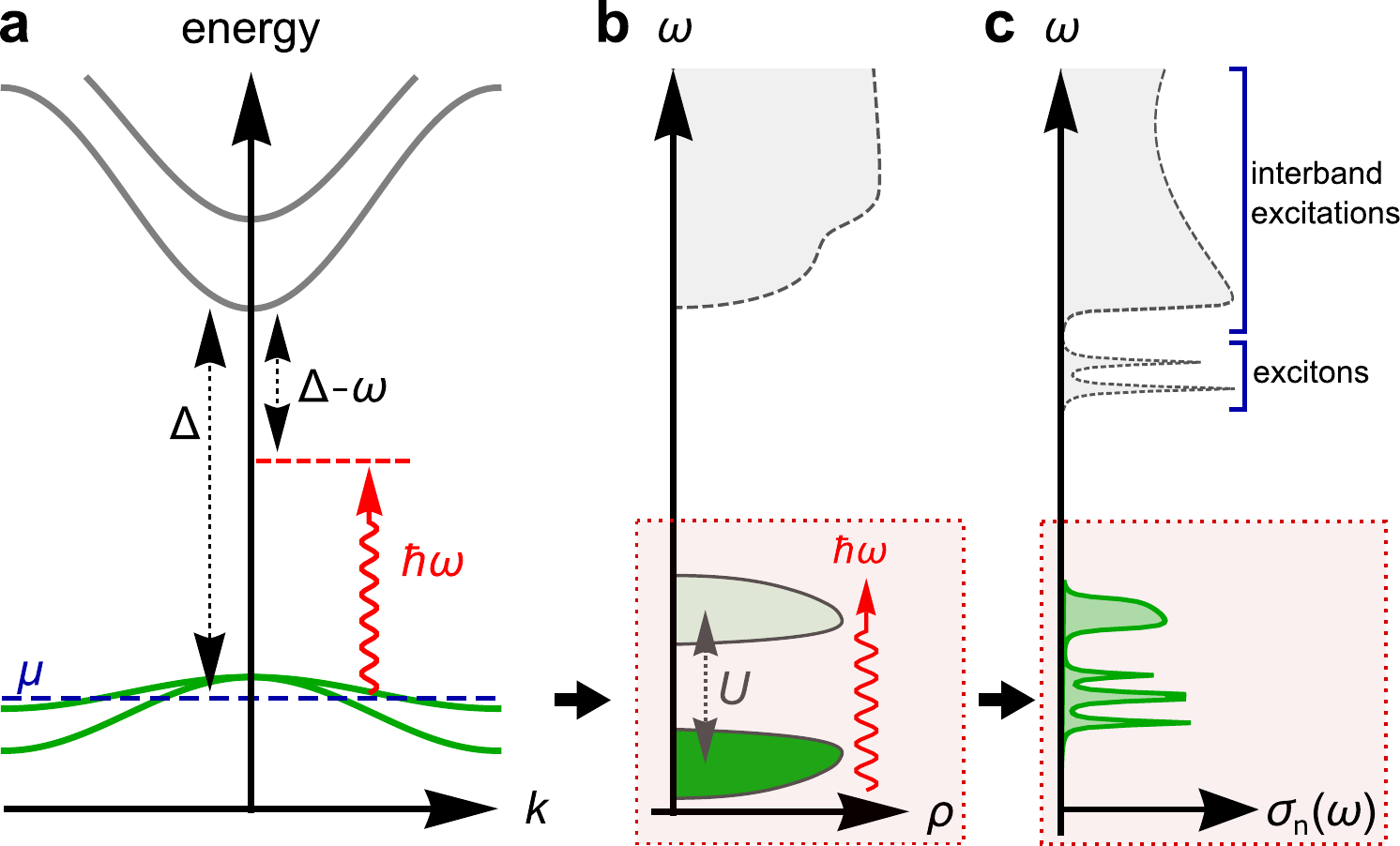}
	\caption{\captiontitle{Coupling Light to Correlated Electrons.} (a) Schematic of a strongly-correlated electron system irradiated with light. If the frequency is detuned from interband transitions $\Delta$ between the partially-filled bands near the Fermi energy and inert higher-energy bands, (b) the light pulse should couple solely to interacting electrons in the material's open shell orbitals, which in turn (c) governs low-frequency non-linear optical responses.}
	\label{fig:schematics}
\end{figure}

\begin{figure*}[t]
	\centering
	\includegraphics[width=0.8\textwidth]{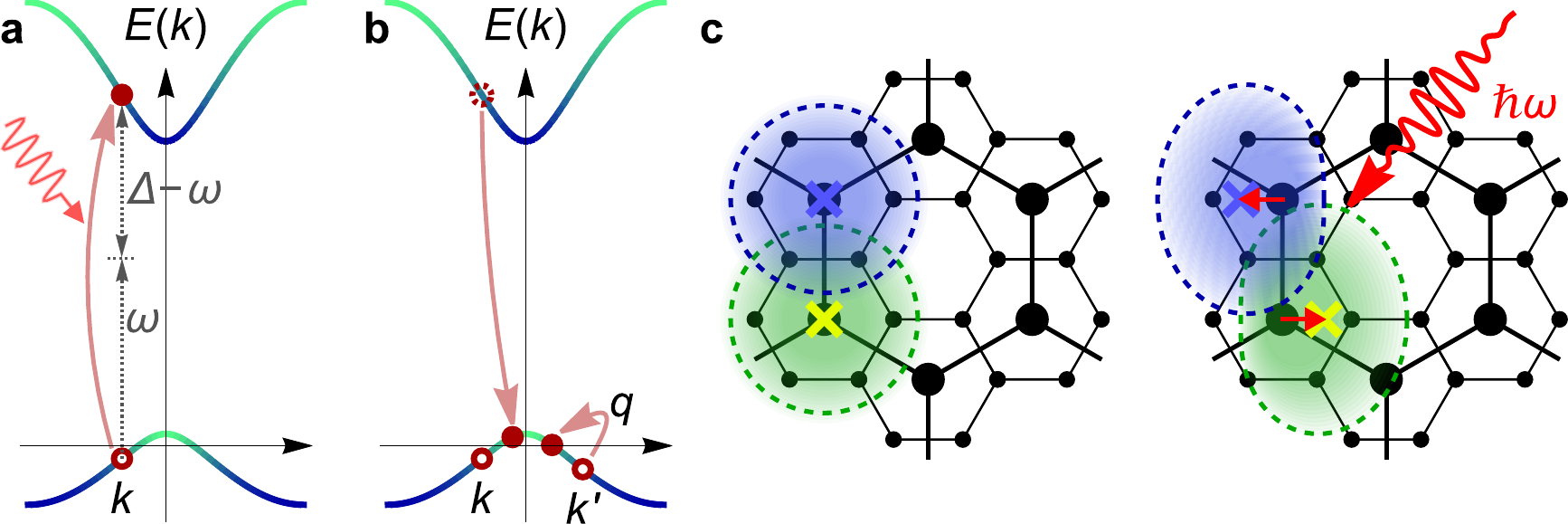}
	\caption{\captiontitle{Quantum geometry and time-dependent Wannier functions.}  \textit{Equilibrium} Bloch bands with rapidly varying orbital character yield interband current matrix elements that diverge with the band gap $\Delta$, which necessitates taking into account highly-detuned processes $\sim \Delta / (\Delta - \omega)$ whereby (a) an electron is photo-excited to an off-resonant inert higher-energy band and subsequently (b) Coulomb scattered off a second electron to create two electron-hole pairs near the Fermi energy. Theses spurious processes cancel precisely when accounting for (c) the concurrent light-induced deformation of both the shape and center of the Wannier orbitals that comprise the low-energy bands, which yields an effective quantum-geometric light-matter interaction.} 
	\label{fig:timedependentWannier}
\end{figure*}

\sectiontitle{Coupling Light to Quantum Matter} In equilibrium, well-established approaches to the many-electron problem in strongly-correlated materials canonically start from low-energy interacting tight-binding descriptions of the open-shell orbitals, constructed either empirically or via calculating their Wannier orbitals \textit{ab initio}. Formally, starting from the Hamiltonian of interacting electrons in a crystal
\begin{align}
	\Ham &= \int d\r ~\hat{\boldsymbol{\Psi}}^\dag \left[ \frac{\hat{\mathbf{p}}^2}{2m} + U(\r) + \frac{\hbar}{4m^2c^2} \hat{\boldsymbol{\sigma}}\cdot \hat{\mathbf{p}} \times \boldsymbol{\nabla}U(\r)
	\right] \hat{\boldsymbol{\Psi}} \notag\\
	&+ \frac{1}{2} \sum_{\sigma\sigma'} \iint d\r d\r' V(\r-\r') \hat{\Psi}_{\sigma}^\dag(\r) \hat{\Psi}_{\sigma'}^\dag(\r') \hat{\Psi}_{\sigma'}(\r') \hat{\Psi}_{\sigma}(\r)  \label{eq:parentHamiltonian}
\end{align}
with crystal potential $U(\r)$, Coulomb interactions $V(\r-\r')$ and spin-orbit coupling, these low-energy models emerge via expressing the interacting problem either in a momentum-space basis of single-particle (Bloch) eigenstates $\mathbf{u}_{n\k}(\r)$ with dispersion $\E_{n\k}$ or, equivalently, a corresponding tight-binding model of maximally-localized Wannier orbitals $\varphi_{m}(\r)$
\begin{align}
	\hat{\boldsymbol{\Psi}}(\r) = \sum_{n\k} e^{i\k\r} \mathbf{u}_{n\k}(\r) ~\C{n\k} = \sum_{m\R} \boldsymbol{\varphi}_{m}(\r - \R) \C{m\R} \label{eq:equilibriumBasis}
\end{align}
and keeping only a small set $n \in \mathcal{A}$ of partially-filled bands or corresponding Wannier orbitals $\{ \boldsymbol{\varphi}_m \}$ that span these bands. Here, $\mathbf{u}$, $\boldsymbol{\varphi}$ include spin, $\R$ is a lattice vector, and $n$ and $m$ index spin-orbit-coupled bands and Wannier orbitals, respectively. Maximally-localized Wannier orbitals in equilibrium
\begin{align}
	\boldsymbol{\varphi}_{m}(\r-\R) = \sum_{\k} e^{i\k(\r-\R)} \tilde{\mathbf{u}}_{m\k}(\r)  \label{eq:equilibriumWannierFunction}
\end{align}
are defined via a gauge-fixing unitary transformation for Bloch functions
\begin{align}
	\tilde{\mathbf{u}}_{m\k}(\r) = \sum_{n} U_{mn}(\k) \mathbf{u}_{n\k}(\r)  \label{eq:gaugeFixedBlochFunctions}
\end{align} 
that minimizes the Wannier spread of $\boldsymbol{\varphi}_{m}$ as a function of the $N$ dimensional unitary matrix $U_{mn}(\k)$ by ``entangling'' a group of $N$ bands near the Fermi energy. Effective Coulomb interactions $V(\r)$ in the resulting tight-binding description are typically of short-ranged Hubbard-Kanamori form due to screening, but can prominently become longer-ranged in systems such as twisted bilayer graphene or fractional topological insulators which have poorly-localized Wannier functions due to geometric or topological obstructions. Furthermore, Wannier functions computed via density functional theory (DFT) approaches must be interpreted as quasiparticle orbitals that already account for a subset of the Coulomb interaction vertex, necessitating in principle the inclusion of functional-dependent ``double counting'' corrections to the effective low-energy theory.

Coupling to light enters the bare Hamiltonian [Eq. (\ref{eq:parentHamiltonian})] via straightforward minimal substitution $\hat{\mathbf{p}} \to \hat{\mathbf{p}} + e\mathbf{A}(t),~ U(\r) \to U(\r) + \phi(\r)$, but describing the resulting coupling of photons to the \textit{effective} electronic degrees of freedom near the Fermi energy remains a profound challenge. To illustrate the insufficiency of conventional approaches for strongly-correlated electron systems and the emergence of a new quantum geometric contribution that encodes the deformation of the Wannier functions with light, we start without loss of generality from a velocity gauge $\mathbf{A}(\r,t) = E_0 \sin(\omega t) / \omega$, $\phi(\r,t) = 0$ description, which yields a diamagnetic $\sim \mathbf{A}^2$ term coupling to the charge density, and a paramagnetic interaction $\hat{\mathbf{J}} \cdot \mathbf{A}$ with current operator matrix elements
\begin{align}
	\mathbf{J}_{nn'}(\k) = \frac{e}{m} \int d\r~ u_{n\k}^\star(\r) ~\frac{\delta H}{\delta \mathbf{A}}~ u_{n'\k}(\r)  \label{eq:currentOperatorMatrixElem}
\end{align}
between bands $n,n'$. These can be separated into intra- and interband contributions $\mathbf{J}_{nn'} = \mathbf{J}^{\textrm{intra}}_{n} \delta_{n,n'} + \mathbf{J}^{\textrm{inter}}_{nn'}$, where the intraband current is given by the band velocity $\mathbf{J}^{\textrm{intra}}_{n} = \frac{e}{\hbar} \boldsymbol{\nabla}_\k \E_{n}(\k)$, and the interband current can be expressed \cite{blount62}
\begin{align}
	\mathbf{J}^{\textrm{inter}}_{nn'}(\k) = i \frac{e}{\hbar}~ \Delta_{nn'}(\k)~ \boldsymbol{\mathcal{A}}_{nn'}(\k)     \label{eq:interbandCurrent}
\end{align}
in terms of the energy gap $\Delta_{nn'}(\k) \equiv \left( \E_{n}(\k) - \E_{n'}(\k) \right)$ between bands and the interband Berry connection $\boldsymbol{\mathcal{A}}_{nn'}(\k) \equiv -i \int d\r~ u_{n\k}^\star(\r) \boldsymbol{\nabla}_\k u_{n'\k}(\r)$, which in turn defines the gauge-invariant non-Abelian Berry curvature $\mathbf{F}^{\mu\nu} = \partial_{k_\mu} \mathcal{A}^{\nu} - \partial_{k_\nu} \mathcal{A}^{\mu} + \left[ \mathcal{A}^{\mu},~ \mathcal{A}^{\nu} \right]$. Notably, the \textit{magnitude} of the interband Berry connection $\left| \boldsymbol{\mathcal{A}}_{nn'}(\k) \right|$ is a gauge-invariant quantity as well, hence cannot be eliminated via a gauge choice for the Bloch states.

Suppose now that the material hosts a set of partially-filled bands at the Fermi energy, schematically depicted in Fig. \ref{fig:schematics}(a) and separated from the nearest filled or empty inert bands via a gap $\Delta$. Physically, if this band gap is larger than the photon frequency $\Delta \gg \omega$, the effective light-induced dynamics should be governed solely by electronic motion within the strongly-interacting partially-filled bands [Fig. \ref{fig:schematics}(b)], with any corrections due to interband transitions becoming suppressed by the inverse detuning $1 / (\Delta - \omega)$ and vanishing rigorously in the large band gap limit $\Delta \to \infty$. Na\"ively, this implies that light couples solely to the intraband velocity $\mathbf{J}^{\textrm{intra}}$ as well as transitions between the partially-filled bands, whereas higher-energy conduction bands or deeper valence bands remain inert. This observation has problematic ramifications, suggesting, for instance, that a Mott insulator forming in a single flat band in a moir\'e heterostructure cannot couple to light, as the band velocity vanishes.

The above intuition fails drastically in a material with non-trivial quantum geometry. If the interband Berry connection $\boldsymbol{\mathcal{A}}$ to an inert band is non-zero, the interband current operator [Eq. (\ref{eq:interbandCurrent})] can immediately be seen to scale with the band gap $\Delta$. Consequently, interband transitions scale as $\Delta / (\Delta - \omega)$ and apparently remain non-negligible even if $\Delta \to \infty$. In free-electron systems such as band insulators, a careful resummation of all interband transitions remains feasible, has been shown to yield correct low-frequency linear and second-order optical responses, and can be altogether circumvented in length gauge calculations. However, this situation changes dramatically in strongly-interacting electron systems, in which the above interband transitions must necessarily become dressed by electron-electron scattering. Fig. \ref{fig:timedependentWannier}(a) schematically depicts the simplest allowed second-order process. Absorption of an off-resonant photon triggers a virtual interband transition ($\sim \mathbf{J}^{\textrm{inter}} \cdot \mathbf{A}$) to an empty high-energy band, followed by interband Coulomb scattering ($\sim U$) off a second electron near the Fermi energy, which returns the photoexcited electron back to the partially-filled band to yield a pair of low-energy electron-hole excitations. The resulting process scales as $\sim U \boldsymbol{\mathcal{A}} \Delta / (\Delta - \omega)$ and crucially again remains nonnegligible even if the interband transition is far from resonance and must be expected to dominate the optical response in strongly-interacting materials, a highly problematic result that na\"ively suggests -- in contradiction with physical intuition -- that high-energy degrees of freedom can contribute to optical responses of correlated materials even if they are far off resonance with the light field. 

\sectiontitle{Quantum-Geometric Gauge Choice for Light-Matter Interactions} A resolution readily follows from observing that the material's equilibrium Bloch states [Eq. (\ref{eq:equilibriumBasis})] and corresponding Wannier functions must \textit{themselves} deform in the presence of the light field [Fig. \ref{fig:timedependentWannier}(b)], in precisely such a manner that this deformation cancels the $\Delta$-divergent contributions [Eq. (\ref{eq:interbandCurrent})] to the interband current matrix elements $\mathbf{J}_{nn'}^{\textrm{inter}}(\k)$ to higher-energy empty conduction states or deeper fully-filled valence states. For an individual isolated band at the Fermi energy, the unique eigenbasis that satisfies this requirement
\begin{align}
	\hat{\boldsymbol{\Psi}}(\r,t) = \sum_{n\k} e^{i\k r} \mathbf{u}_{n,\k+\mathbf{A}(t)}(\r) ~ e^{-i \phi_n(\k,t)}~ \C{n\k}  \label{eq:geometricFieldSingleBand}
\end{align}
admits an appealingly-simple physical interpretation as Bloch states for the \textit{gauge-invariant} momenta subjected to a quantum-geometric phase
\begin{align}
	 \phi_n(\k,t) = \hspace{-0.2cm}\int\limits_{\k}^{\k+\mathbf{A}(t)} \hspace{-0.3cm} \boldsymbol{\mathcal{A}}_{n}(\k') \cdot d\k'  \label{eq:geometricPhase}
\end{align}
that encodes their parallel transport back to the gauge-dependent momentum $\k$, with $\Aberry_{n}(\k)$ the band's (Abelian) Berry connection. The residual gauge freedom of the equilibrium Bloch state $\mathbf{u}_{n,\k}$ should be fixed to guarantee a maximally-localized Wannier function after Fourier transform, as discussed below. Notably, the resulting theory remains formulated -- without loss of generality -- in velocity gauge $\Agauge \neq 0, \phi = 0$ for the electromagnetic field, but amounts to a new gauge choice for the Bloch wave functions by virtue of the quantum-geometric phase factor.

Applied to the bare Hamiltonian (\ref{eq:parentHamiltonian}) with $\Ham - \int d\r~ \hat{\boldsymbol{\Psi}}^\dag(\r,t) i\partial_t \hat{\boldsymbol{\Psi}}(\r,t)$, one immediately finds that the resulting interband light-matter coupling to inert bands $\sum_{n \neq n',\k} \Aberry_{n,n'}(\k + \Agauge(t)) \cdot \mathbf{E}(t)~ \CD{n\k} \C{n'\k}$ remains independent of the band gaps $\Delta$, guaranteeing that interband transitions to inert bands are suppressed by the detuning $\sim 1/(\Delta - \omega)$ as physically expected. Crucially however, and in addition to the usual Peierls substitution, the effective light-matter-coupled Hamiltonian for the partially-filled band (dropping its index $n$)
\begin{align}
	\Ham &= \sum_{\k} \E(\k + \Agauge(t))~ \CD{\k} \C{\k} \notag\\
		&+ \frac{1}{L} \sum_{\k\k'\q} V_{\k\k'\q}^{\rm eff}(\Agauge(t))~ \CD{\k+\q} \CD{\k'-\q} \C{\k'} \C{\k}  \label{eq:singleBandEffectiveHamiltonian}
\end{align}
now acquires a field-dressed time-dependent interaction
\begin{align}
	V_{\k\k'\q}^{\rm eff} &= V(\q) \braket{u_{\k+\q+\Agauge(t)}}{u_{\k+\Agauge(t)}} \braket{u_{\k'-\q+\Agauge(t)}}{u_{\k'+\Agauge(t)}} \notag\\ &\times~ e^{i \left[ \phi(\k+\q,t) + \phi(\k'-\q,t) - \phi(\k',t) - \phi(\k,t) \right]  }  \label{eq:blochInteractionVertex}
\end{align}
which describes photons directly dressing the band's effective electron-electron interactions and encodes the Coulomb interaction assisted absorption/emission processes of Fig. \ref{fig:timedependentWannier}(a).

The generalization to multiband models follows straightforwardly via promoting Eq. (\ref{eq:geometricPhase}) to a non-Abelian geometric phase. Given a set of $N$ ``entangled'' bands $\mathcal{G}$ near the Fermi energy, the light-induced geometric phase factor $e^{i \phi_n(\k,t)}$ must now be replaced by an $N$-dimensional unitary rotation
\begin{align}
	\mathbf{Q}(\k, t) = \hat{\mathcal{P}}~ \exp\left\{ -i \hspace{-0.2cm} \int\limits_{\k}^{\k+\Agauge(t)} \hspace{-0.3cm} \Aberry_{\mathcal{G}}(\k') \cdot d\k' \right\}  \label{eq:nonAbelianGeometricPhase}
\end{align}
where $\hat{\Aberry}_{\mathcal{G}}$ is the $N$-band non-Abelian Berry connection and $\hat{\mathcal{P}}$ denotes the momentum space path ordering operator. $\mathbf{Q}(\k, t)$ describes the parallel transport of all $N$ states from the gauge-invariant momentum back to $\k$, thereby admixing the original $N$ bands [see Appendix \ref{appendix:lightmattercoupling}].

\sectiontitle{Time-Dependent Wannier Functions} Insight into the origin of the field-dressed interaction can be gleaned from a Fourier transform to real space $\C{m\R} = \frac{1}{\sqrt{L}} \sum_{\k} e^{i\k\R} \C{m\k}$. The resulting field operator
\begin{align}
	\hat{\boldsymbol{\Psi}}(\r,t) = \sum_{m\R} e^{-i \Agauge(t) \cdot (\r-\R)} \boldsymbol{\varphi}_{m\R}(\r,t) ~\C{m\R}
\end{align}
can now be written in terms of products of Peierls phases and $N$ photon-dressed Wannier functions
\begin{align}
	\boldsymbol{\varphi}_{m\R}(\r, t) &= \sum_{\k nn'} e^{i\k(\r-\R)} \left[ \hat{\mathcal{P}}~ e^{i \hspace{-0.3cm} \int\limits_{\k}^{\k-\Agauge(t)} \hspace{-0.3cm} \Aberry_{\mathcal{G}}(\k') \cdot d\k' } \right]_{n'm} \hspace{-0.35cm} U_{nn'}^\k \mathbf{u}_{n\k}(\r)   \label{eq:timeDependentWannierMultiband}
\end{align}
These \textit{time-dependent Wannier functions} constitute a central result of this work. They provide an effective description of the coupling of photons to strongly-interacting electrons by accounting for the light-induced deformation of the effective electronic orbitals $m$ that compose the $N$ partially-filled bands near the Fermi energy. Their Bloch states $\mathbf{u}_{n\k}(\r)$ are gauge fixed via an $N$ dimensional unitary transformation $U_{mn}^\k$ to yield maximally-localized Wannier functions in equilibrium, and subjected to a field-dependent unitary transform -- the quantum-geometric non-Abelian phase -- which changes the Wannier function localization as a function of time. Here, the non-Abelian Berry connection $\Aberry^{mm'}_{\mathcal{G}}(\k) = \int d\r  \sum_{nn'} U^\star_{nm}(\k) \mathbf{u}_{n\k}(\r) (-i\boldsymbol{\nabla}_\k) \mathbf{u}_{n'\k}(\r) U_{n'm'}(\k)$ must be defined with respect to the gauge-fixed basis.

The resulting orbital changes both its shape and its Wannier center $\left<\r \right>_m$ as a function of the field as depicted schematically in Fig. \ref{fig:timedependentWannier}(c), with both effects remarkably depending solely on the quantum geometry of the band. The total ``center of mass'' Wannier center of all $N$ orbitals per unit cell however remains invariant. Different path choices for the parallel-transport coefficients constitutes a residual gauge freedom of the time-dependent Wannier construction and are unitarily equivalent, with a change of paths amounting to a transformation using Wilson loop operators for $\Aberry_{\mathcal{G}}$; for weak fields, changes to the time-dependent localization remain negligible for smooth and short-distance paths. Notably, the formalism remains well-defined only for non-singular Berry connections and breaks down for a Chern band. Conversely, coupling light to correlated electrons in symmetry-protected topological bands can be captured faithfully by choosing exponentially localized Wannier representations that implement the protecting symmetries in a non-local manner \cite{soluyanov10}, as will be demonstrated for twisted bilayer graphene.

Notably, in contrast to previous approaches, the time dependent Wannier basis permits a consistent velocity gauge formulation of the light-driven dynamics of interacting electrons, whereby electrons couple solely to the magnetic vector potential $\Agauge$, with $\phi = 0$. The formulation is asymptotically exact if resonances to bands not included in the tight binding description are sufficiently detuned. However, the choice of velocity gauge is arbitrary, in principle, and the resulting theory is gauge invariant. For instance, a length gauge description readily follows from a Power-Zienau-Wooley transformation in the time-dependent Wannier basis $\C{m\mathbf{R}} \longrightarrow \C{m\mathbf{R}} e^{i \R \cdot \Agauge(t)}$, which eliminates Peierls factors in favor of a dipole contribution $\Ham_{\rm dip} = e \sum_{\R} \left( \mathbf{E}(t) \cdot \R \right) ~ \CD{m\R} \C{m\R}$ that couples to the electric field $\mathbf{E}$, in addition to the motion of Wannier orbitals. Alternatively, the latter can be formally transformed via $\C{m\mathbf{R}} \longrightarrow \sum_{\R'} \C{m\mathbf{\R'}} e^{i \R'\cdot\Agauge(t)} \left( \sum_\k e^{i\k(\R-\R')} \hat{\mathcal{P}} \exp\{ i \int_{\k}^{\k+\Agauge} d\Aberry \} \right)$ and yields theory of non-local dipole transitions $\sum_{\R,\R'} (D_{\R,\R'} + \R ) \cdot \mathbf{E}(t)$ \cite{golez19} \textit{in addition} to non-local interactions. Here, the concurrent necessity to account for non-local dipole transitions and non-local interactions is a key constraint due to quantum geometric bounds discussed below \cite{roy14} and can be viewed as an alternate cause of the responses described in this work. We note that dipole terms are formally defined only in a finite system and have norms that diverge with system size, both of which makes velocity-gauge formulations preferable for analytical and computational treatments of the light-matter-coupled many-body problem.

\begin{figure*}[t]
	\centering
	\includegraphics[width=0.85\textwidth]{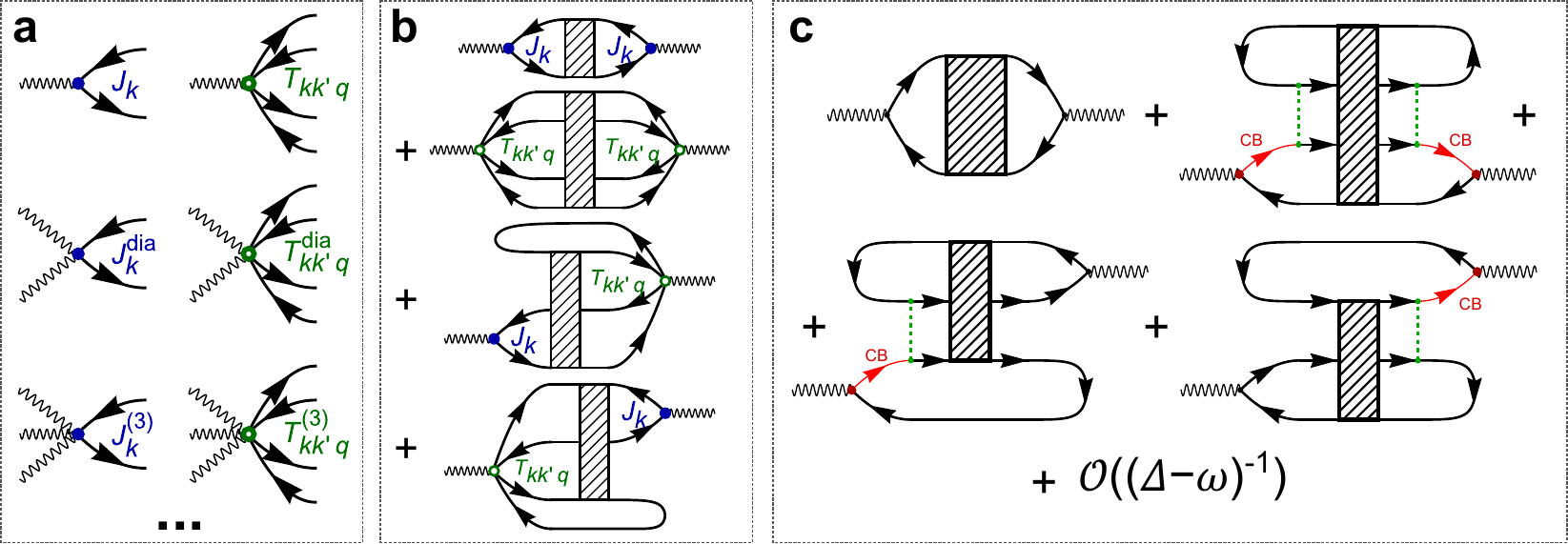}
	\caption{\captiontitle{Spectroscopy of time-dependent Wannier functions.} (a) Non-linear optical responses are jointly governed by single-particle (left) and interaction-assisted (right) current vertices that emerge from time-dependent deformations of the Wannier orbitals, with the latter dominating the response for strongly-correlated electrons. Consequently, already (b) linear optical absorption comprises diagrammatic contributions with pairs of single-particle ($J$) and two-particle ($T$) current vertices as well as mixed terms.  Shaded boxed denote fully dressed two-, three- and four-particle propagators that resum all self energy contributions from intraband Coulomb interactions. (c) Analogous velocity-gauge calculations in the original Bloch band basis must account for Coulomb assisted interband transitions [depicted in Fig. \ref{fig:timedependentWannier}(a)] via an intermediate conduction band (or deeper valence band) state CB. Green dashed lines denote \textit{interband} Coulomb scattering that scatters an electron in an inert band off a valence electron, with both final states located in the valence bands of interest.}
	\label{fig:diagrams}
\end{figure*}

Starting from the usual equilibrium Bloch Hamiltonian for Wannier orbitals, which is formally defined
\begin{align}
	h_\k^{nn'} = \sum_{l} U^\star_{ln}(\k) \E_{l}(\k) U_{ln'}(\k)
\end{align}
via the band dispersion $\E_{l}(\k)$ of Bloch state $l$ and the Wannier gauge unitary transformation $U_{ln}(\k)$, the photon-dressed kinetic hopping matrix elements
\begin{align}
	t_{\R-\R'}^{mm'} = \sum_{\k nn'} e^{i\k(\R-\R')} Q^\star_{nm}(\k-\Agauge,t) h_\k^{nn'} Q_{nm'}(\k-\Agauge,t)
\end{align}
remain invariant for a single isolated band, for which $\mathbf{Q}$ is a scalar phase, but admix the local orbitals as a function of the driving field in multiband descriptions. The resulting lattice Hamiltonian
\begin{align}
	\Ham &= \sum_{\R\R'} \sum_{mm'} t^{mm'}_{\R-\R'}(t) ~ e^{i (\R-\R') \cdot \Agauge(t)}~ \CD{m\R} \C{m\R'} ~+ \notag\\
		&+ \sum_{\substack{ \R_1\cdots\R_4 \\ m_1\cdots m_4} } V^{m_1 \dots m_4}_{\R_1 \dots \R_4}(t)~ e^{i (\R_1+\R_2-\R_3-\R_4) \cdot \Agauge(t)} ~\times \notag\\
		&~~~~~~~~~~~~~~~~~~~~~~~\times \CD{m_1\R_1} \CD{m_2\R_2} \C{m_3\R_3} \C{m_4\R_4}
\end{align}
includes Peierls phase factors by virtue of working in velocity gauge. Crucially, the field-dressed electron-electron interactions can now be understood to emerge from two distinct contributions: (1) a time-dependent modification of the Coulomb integrals
\begin{align}
	V^{m_1 \dots m_4}_{\R_1\dots\R_4}(t) = \iint d\r d\r' &V(\r-\r') \left[ \varphi^\star_{m_1\R_1}(\r, t) \varphi^\star_{m_2\R_2}(\r', t) \right. \notag\\
		\times & \left.  \varphi_{m_3\R_3}(\r',t) \varphi_{m_4\R_4}(\r,t) \vphantom{\varphi^\star_{\R_1}} \right]
\end{align}
due to quantum-geometric deformations of the Wannier functions, and (2) a Peierls phase acquired by \textit{non-local} scattering processes with $\R_1 + \R_2 - \R_3 - \R_4 \neq 0$ that describe correlated hopping or spin exchange processes.

Importantly, these two effects are not independent, as both are rooted in the quantum geometry of the bands. Non-local two body scattering (or, equivalently, the $\k,\k'$ dependence of the momentum space interaction vertex $V(\k,\k',\q)$ of Eq. (\ref{eq:blochInteractionVertex})) is bounded from below by the Wannier function spread functional $\left< \r^2 \right> - \left< \r \right>^2$, which in turn is bounded by the trace of the gauge invariant Fubini-Study metric \cite{marzari11}. For a single band, the metric reads $g^{\mu\nu}(\k) = \frac{1}{2} \bra{\partial_{k_\mu} \tilde{u}_{\k}} [ 1 - \ketbra{u_\k}{u_\k} ] \ket{\partial_{k_\nu}} + (\mu \leftrightarrow \nu)$. The Fubini-Study metric is bounded from below $\textrm{tr}~ \mathbf{g}(\k) \geq |\Omega(\k)|$
by the band's Berry curvature $\Omega(\k) = \epsilon_{\mu\nu} \partial_{k_\mu} \mathcal{A}^\nu(\k)$  \cite{roy14} which in turn determines (for a smooth maximally-localized gauge choice) the non-Abelian Berry connection $\mathcal{A}$ which governs the time-dependent deformation of the Wannier orbitals.

\sectiontitle{Linear and Nonlinear Optical Response}
These observations are prominently reflected in calculations of nonlinear optical responses. In velocity gauge, the time-dependent Wannier basis readily defines paramagnetic, diamagnetic and higher-order current operators that contribute to non-linear responses such as second harmonic generation or shift currents, by varying $\Ham$ with respect to the gauge field
\begin{align}
	\hat{J}_\mu^{\textrm{para}} = \frac{\delta \Ham}{\delta A^\mu},~ \hat{J}_{\mu\nu}^{\textrm{dia}} = \frac{\delta^2 \Ham}{\delta A^\mu \delta A^\nu},~  \hat{J}_{\mu\nu\sigma}^{(3)} = \frac{\delta^3 \Ham}{\delta A^\mu \delta A^\nu \delta A^\sigma}
\end{align}
Crucially, each current operator now entails both a single-particle and two-particle contribution
\begin{align}
	\hat{J}_\mu &= \sum_{\k mm'} J_{\k\mu}^{mm'} \CD{m\k} \C{m'\k} \notag\\
    &+ \sum_{\substack{\k\k'\q \\ m_{1\dots 4}}} T_{\k\k'\q\mu}^{m_{1\dots 4}} \CD{m_1\k+\q} \CD{m_2\k'-\q} \C{m_3 \k'} \C{m_4 \k}
\end{align}
reflecting that both kinetic hoppings and Coulomb interactions are dressed by photons. For the paramagnetic current operator, the single-particle ($\left[ \mathbf{J}_\mu^{\rm para} \right]_{mm'}$) contribution
\begin{align}
	\mathbf{J}^{\rm para}_{\k\mu} = \frac{\partial \mathbf{h}_\k}{\partial k_\mu} + i \left[ \Aberry^\mu_{\k},~ \mathbf{h}_\k \right]
\end{align}
includes both the usual $\k$ derivative of the Bloch Hamiltonian $\mathbf{h}_\k$ due to the Peierls phase and a commutator of $\mathbf{h}_\k$ and the non-Abelian Berry connection $\Aberry$ for the $N$ bands of interest which accounts for the deformation of the Wannier orbitals. Similarly, the two-particle paramagnetic vertex
\begin{align}
	\mathbf{T}^{m_1 \dots m_4}_{\k\k'\q\mu} &= 2 V(\q) ~\rho^{m_2m_3}_{\k'-\q,\k'}  \notag\\ 
	\times &\left[ \frac{\partial \rho_{\k+\q,\k} }{\partial k_\mu} +  i \left(\Aberry^\mu_{\k+\q} \boldsymbol{\rho}_{\k+\q,\k} - \Aberry^\mu_{\k} \boldsymbol{\rho}_{\k+\q,\k} \right) \right]^{m_1m_4}  \label{eq:twoparticleCurrentVertex}
\end{align}
can be defined in terms of the projected density operator
\begin{align}
	\rho_{\k,\k'}^{mm'} = \braket{u_{m\k}}{u_{m'\k'}}
\end{align}
and includes both Peierls phase contributions (non-vanishing $\k,\k'$ derivatives of $\rho_{\k,\k'}$ due to non-local two-particle scattering) and a $\Aberry_\mathcal{G}^\mu$ contribution due to the light-induced deformations of the Wannier orbital. The $i^{\rm th}$ order current operator can be defined using a covariant derivative
\begin{align}
	D_\mu \mathbf{F} = \partial_{k_\mu} \mathbf{F} + i \left[ \Aberry^{\mu}_{\mathcal{G}},~ \mathbf{F} \right]
\end{align}
which acts on $\mathbf{h}_\k$ and $\boldsymbol{\rho}_{\k+\q,\k} \otimes \boldsymbol{\rho}_{\k'-\q,\k'}$ to yield single-particle ($J$) and two-particle ($T$) contributions respectively. Here, care must be taken to interpret the commutator $[ \Aberry^{\mu},~ \boldsymbol{\rho}_{\k,\k'}] = \Aberry^{\mu}_{\k} \boldsymbol{\rho}_{\k,\k'} - \boldsymbol{\rho}_{\k,\k'} \Aberry^{\mu}_{\k'}$.

Fig. \ref{fig:diagrams}(a) depicts these processes diagrammatically, in terms of two types of $i$-photon vertices that scatter either a single electron or a pair of electrons \textit{within} the bands near the Fermi energy. For simplicity, consider first the ramifications for linear optical conductivity $\sigma(\omega)$. For a single isolated, dispersionless and topologically trivial band, the real part of $\sigma(\omega)$ was shown to remain finite at low-frequencies due to Coulomb interactions despite a vanishing band velocity \cite{mao22}. Such a contribution can be immediately seen to arise from perturbations of the two-particle Coulomb interactions in the time-dependent Wannier basis to linear order in the field. It formally involves a pair of two-particle-scattering photon vertices $T_{\k\k'\q}$ with a fully-dressed four-particle propagator for the partially-filled flat band [Fig. \ref{fig:diagrams}(b), second diagram]. More generally, the linear optical response for a group $\mathcal{G}$ of dispersive partially-filled bands instead mirrors thermal conductivity calculations, with $N$-band paramagnetic and diamagnetic current vertices comprising both a single-particle and two-particle contribution and requiring the evaluation of a two-particle, a four-particle and two mixed diagrams as shown in Fig. \ref{fig:diagrams}(b). The resulting multiband current operators obey a reduced $f$-sum rule for the low-energy bands of the model $\Real \int_0^\infty \sigma(\omega) d\omega  = \pi \langle \hat{J}^{\rm dia} \rangle$. 

To illustrate how these Coulomb interaction mediated contributions to $\sigma(\omega)$ and higher-order responses emerge from apparent interband transitions in the original Bloch basis [Fig. \ref{fig:timedependentWannier}(a)], consider the corresponding calculation in the usual velocity gauge for the equilibrium bands, depicted in Fig. \ref{fig:diagrams}(c). Here, the divergence of the interband current operators [Eq. (\ref{eq:interbandCurrent})] with the band gap $\Delta$ necessitates that, for low-frequency photons (far detuned from interband transitions), \textit{any diagram} that involves an \textit{equal number} of interband transitions (current vertices, $\sim \Aberry \Delta$) and propagators for high-energy inert bands (red arrows, $\sim 1/(\Delta-\omega)$) must contribute $\sim \mathcal{O}(1)$ to the optical response, with subleading contributions suppressed by the inverse detuning. Already for optical absorption, this necessitates carefully resumming interaction-mediated interband scattering processes over \textit{all bands} of the solid, a herculean task. Importantly however, the current vertices and diagrams in the time-dependent Wannier basis [Fig. \ref{fig:diagrams}(a), (b)] can now be readily seen to emerge from neglecting the internal dynamics of the propagator $1 / (\omega - \E_{\k}^{\rm CB} - \Delta)$ for the virtual photoexcited electron in the inert conduction band (CB), which is valid if the detuning to interband transitions is large; in this case, the propagator for inert bands can be contracted and all virtual interband contributions can be resummed to again yield the effective two-particle current vertex [Eq. \ref{eq:twoparticleCurrentVertex}] that depends solely on the quantum geometry of the partially-filled bands near the Fermi energy [see Appendix \ref{appendix:diagrams}]. The correspondence between conventional but careful response calculations in velocity gauge and optical responses from the effective time dependence of Wannier functions is mandatory due to gauge invariance and reveals the central advantage of the quantum-geometric gauge as a faithful effective theory of low-frequency light matter interactions which moreover readily encompasses non-perturbative driving and dynamics as discussed below.

\sectiontitle{Quantum Geometric Optical Response in a Strongly-Interacting Topological Hubbard Chain}

\begin{figure*}[t]
	\centering
	\includegraphics[width=0.9\linewidth]{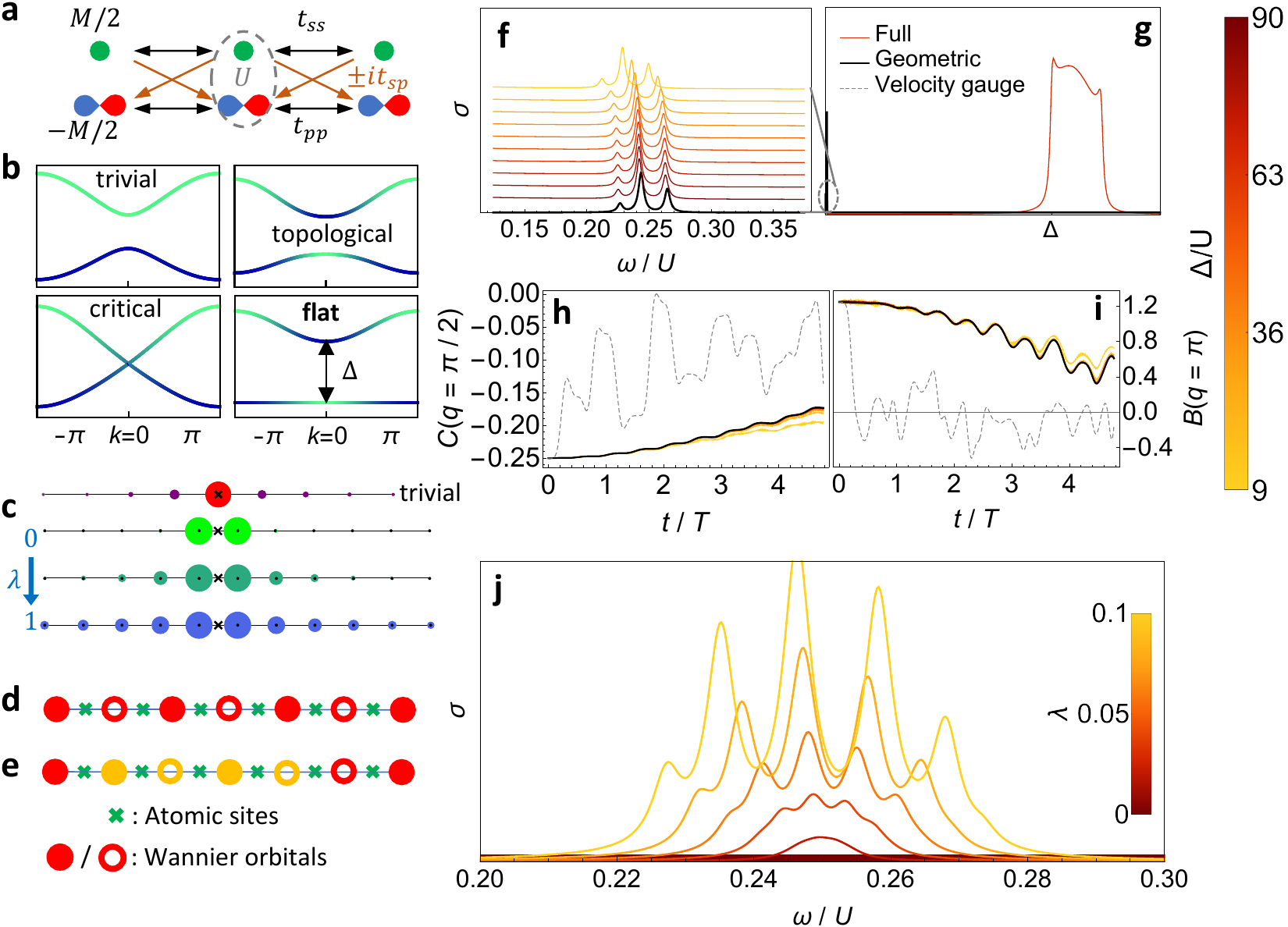}
	\caption{\captiontitle{Spectroscopy and dynamics of an interacting atomically-obstructed chain.} (a) Schematic of an interacting two-orbital chain, which hosts (b) an exact flat band and a tunable topological band inversion distinguished by maximally-localized Wannier orbitals that are centered (c) on atomic sites or bonds and exhibit an independently tunable extent.  (d) and (e) schematically depict the bond density wave ground state at quarter filling, and its elementary domain wall excitations. (f), (g), depict low-frequency and high-frequency (interband) optical absorption computed from the quantum-geometric low-energy theory of the coupling of light and Wannier orbitals, and for the full two-band model of atomic orbitals as a function of the band gap $\Delta$, revealing the geometric origin of the low-frequency response up to interband corrections that vanish with increasing detuning $\Delta - \omega \to \infty$. (h) and (i) 
 demonstrate analogous behavior for time-dependent charge $C(q)$ and bond $B(q)$ correlation functions upon irradiation with low-frequency light. The low-frequency optical response (j) reveals an emergent domain wall mode which exhibits both a dispersion and coupling to light that depends Wannier orbital extent $\lambda$, providing a spectroscopic probe of the system's geometrical obstruction.}
	\label{fig:toymodel}
\end{figure*}

As a first illustration of ramifications of the quantum geometric coupling of light to Wannier functions in correlated materials, consider an interacting two-band Hubbard chain, depicted schematically in Fig. \ref{fig:toymodel}(a), with Hamiltonian
\begin{align}
    \Ham &= \sum_{k\alpha\beta} h_{\alpha\beta}(k)~ \CD{k\alpha} \C{k\beta} + U \sum_i \ND{i,s} \ND{i,p}  \label{eq:HTwoBandChain}
\end{align}
with Coulomb repulsion $U$ between two orbitals $\alpha = s, p_x$ per site $i$ and a Bloch Hamiltonian
\begin{align}
    \mathbf{h}(k) &= \frac{M}{2} ~\hat{\sigma}_z - (t_{ss}+t_{pp})\cos(k) ~\hat{\sigma}_z \notag\\
                  &- (t_{ss}-t_{pp})\cos(k) ~\hat{\sigma}_0 + 2t_{sp} \sin(k)  ~\hat{\sigma}_y
\end{align}
where $M$ is a local orbital splitting, $t_{\alpha\beta}$ denotes hoppings between orbital $\alpha$ and $\beta$, and $\sigma_\nu$ are orbital Pauli matrices. As the bare model is composed of atomic orbitals, coupling to light can be approximated without loss of generality via a Peierls substitution $k \to k + A(t)$ \footnote{Local dipole transitions can be safely neglected by assuming that the three-dimensional atomic orbitals are of $p_x$, $p_z$ type.}.

The inversion-symmetric model exhibits a topological $\mathbb{Z}_2$ band inversion to an obstructed atomic insulator as $M/2-t_{ss}-t_{pp}$ switches sign [Fig. \ref{fig:toymodel}(b)], for which the Wannier centers shift from atomic sites to bonds. Moreover, the model admits an exact flat band if $t_{sp} = \sqrt{\left(t_{ss}+t_{pp}\right)^2 / 4 - (M/4)^2}$ is satisfied. To characterize its behavior, it is convenient to define three \textit{independent} band structure parameters $\epsilon_0 = (t_{ss} - t_{pp})/2 - M/4$, $\Delta = 2(t_{ss} + t_{pp}) - M$ and $\lambda = -M / [2 (t_{ss} + t_{pp})]$. Here, $\epsilon_0$ controls the dispersion $-2\epsilon_0 \cos(k)$ of the lower band and will henceforth be set to zero, $\Delta$ is the band gap to an inert conduction band, and $\lambda$ controls the spread of the maximally-localized Wannier function for the lower band with gauged Bloch state $\ket{u(k)} = [ \sqrt{1+\lambda} (e^{-ik} + 1),~ \sqrt{1-\lambda} (e^{-ik} - 1) ]^\top / N_k$ with $N_k = 2\sqrt{1 + \lambda \cos(k)}$. The Wannier function is perfectly localized on a pair of neighboring sites for $\lambda = 0$, and has increasing spread with an exponential tail as $|\lambda|$ increases [see Fig. \ref{fig:toymodel}(c) and Appendix \ref{appendix:1Dchaintoymodel}].

The confluence of electronic interactions, tunable atomically-obstructed Wannier orbitals, and a rigorous flat band limit with vanishing single-electron velocity establishes the topological two-band Hubbard chain as an ideal illustration of the quantum-geometric coupling of light and correlated electrons. Suppose now that the lower flat band is half-filled (overall quarter filling). Remarkably, its ground state admits an exact solution as a $q = \pi$ bond density wave (BDW) [see Appendix \ref{appendix:exactCDWsolution}] which is schematized in Fig. \ref{fig:toymodel}(d).

We first establish the validity of the quantum-geometric gauge by studying optical absorption $\textrm{Re}~ \sigma(\omega)$ using large-scale exact diagonalization. Here, while the large frequency behavior for $\omega > \Delta$ is governed by single-particle interband transitions [Fig. \ref{fig:toymodel}(g)], the low-frequency \textit{intraband} response $\omega < \Delta$ is dictated entirely by the coupling of light to Wannier orbitals as the single-electron velocity vanishes. Starting from the introduced \textit{effective} quantum-geometric light-matter Hamiltonian for the single partially-filled band [Eqs. (\ref{eq:singleBandEffectiveHamiltonian}), (\ref{eq:blochInteractionVertex})], Fig. \ref{fig:toymodel}(f) compares the ``geometric'' optical conductivity $\textrm{Re}~ \sigma(\omega < \Delta)$ (black line) with a direct computation of $\sigma(\omega)$ for the full two-band model as a function of the band gap $\Delta$ and for a chain length $L=14$. As anticipated, the quantum-geometric response that arises purely from deformations of the Wannier orbitals mirrors precisely the exact response at low-frequencies, up to an energetic shift from residual detuned interband transitions that vanishes exactly as the detuning to the inert band becomes large with $\Delta \gg \omega$.

Similar observations apply to the full non-equilibrium dynamics of charge $C(q) = \sum_{q} \langle \hat{\rho}_{q} \hat{\rho}_{-q} \rangle$ and bond $B(q) = \sum_{q} \langle \hat{\Delta}_q \hat{\Delta}_{-q} \rangle$ correlation functions upon irradiation with a pump field $A(t) = A_0 \sin(\omega t)$ with $\omega = U/4$ and $A_0 = 1$ in dimensionless units, depicted in Fig. \ref{fig:toymodel}(h) and (i). Here, $\rho_{q}$ is the density operator and $\hat{\Delta}_q = \sum_{k\alpha\beta} (e^{i(k+q)}+(-1)^\alpha)(e^{-ik}+(-1)^\beta)  \CD{k+q, \alpha}\C{k,\beta}$ describes bond-localized charges. The strong pump field is resonant with BDW excitations and results in the slow melting of bond density correlations, with the dynamics again exactly reproduced in the quantum-geometric theory in the limit $\Delta / \omega \to \infty$. In comparison, a na\"ive conventional velocity gauge calculation in the single band (dashed line) fails to even qualitatively reproduce the emergent dynamics, a consequence of the poor localization of the Wannier functions as expected.

Strikingly, closer inspection of dependence of the low-frequency response on the Wannier spread $\lambda$ shown in Fig. \ref{fig:toymodel}(j) reveals an emergent collective domain wall pair excitation, the dynamics of which is itself entirely geometric in the flat band limit and dictated by the bound on the spread of Wannier functions. The mode is depicted schematically in Fig. \ref{fig:toymodel}(e). Its coupling to light depends on the Wannier function extent and vanishes in the limit of a strictly-local orbital $\lambda = 0$ while growing in magnitude as $\lambda \neq 0$, coinciding with a broadened mode dispersion. Insight into the latter can be gleaned from expanding the low-energy interacting Wannier Hamiltonian for small $\lambda$
\begin{align}
	\Ham &= \frac{U}{8} \sum_{i} \left[ \left( \ND{i} + \ND{i+1} \right)^2 + \frac{\lambda}{2} \CD{i+1} \C{i}\left( \ND{i+2} + \ND{i-1} \right) + \hc \right]\label{eq:perturbedHamiltonianU} 
\end{align}
with $\C{i}$ the fermionic operators for Wannier orbitals. For $\lambda = 0$, the BDW domain wall pair excitation shown in Fig. \ref{fig:toymodel}(e) is an eigenstate with fixed interaction energy $U/4$ and cannot move. However, as $\lambda$ deviates from zero, the finite extent of the bond-centered Wannier orbitals yields subdominant correlated hopping processes that permit the domain wall pair to move, which furnishes the mode with a finite dispersion observed in Fig. \ref{fig:toymodel}(j) that thus serves as a spectroscopic probe of the system's quantum geometry and Wannier extent.

\begin{figure*}[t]
	\centering
	\includegraphics[width=\textwidth]{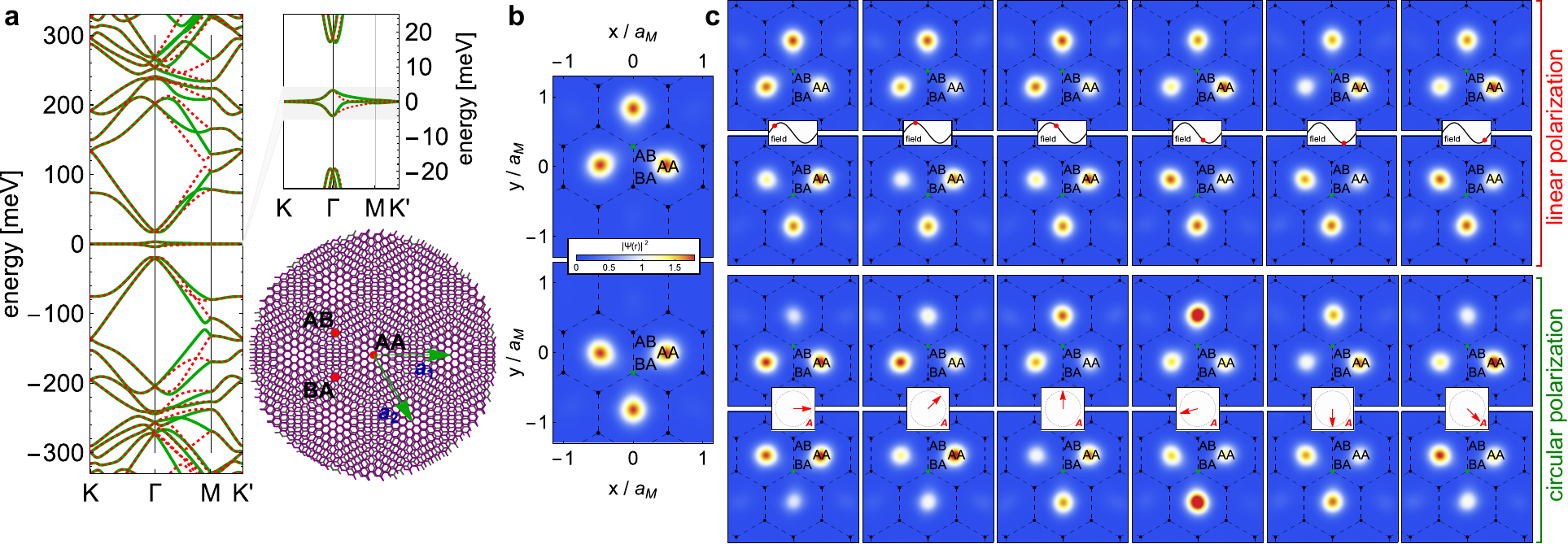}
	\caption{\captiontitle{Photon-dressed Wannier functions in twisted bilayer graphene.} (a) Band structure and moir\'e lattice of twisted bilayer graphene around the magic angle $\theta \sim 1.08^\circ$. An almost-dispersionless band (inset) emerges at charge neutrality, with the charge density accumulating in the AA regions of the moir\'e lattice. (b) ``Fidget-spinner'' Wannier functions of the twisted bilayer graphene flat band in equilibrium are centered at the AB/BA sites but peaked on the three neighboring AA regions. (c) Snapshots of the photon-dressed time-dependent Wannier functions for linearly (top row) and circularly (bottom row) polarized THz pulses. Insets depict the instantaneous electric field.}
    \label{fig:TBG}
\end{figure*}

\sectiontitle{Steering Correlated Phases in Twisted Bilayer Graphene with Light}

\begin{figure*}[t]
	\centering
	\includegraphics[width=\textwidth]{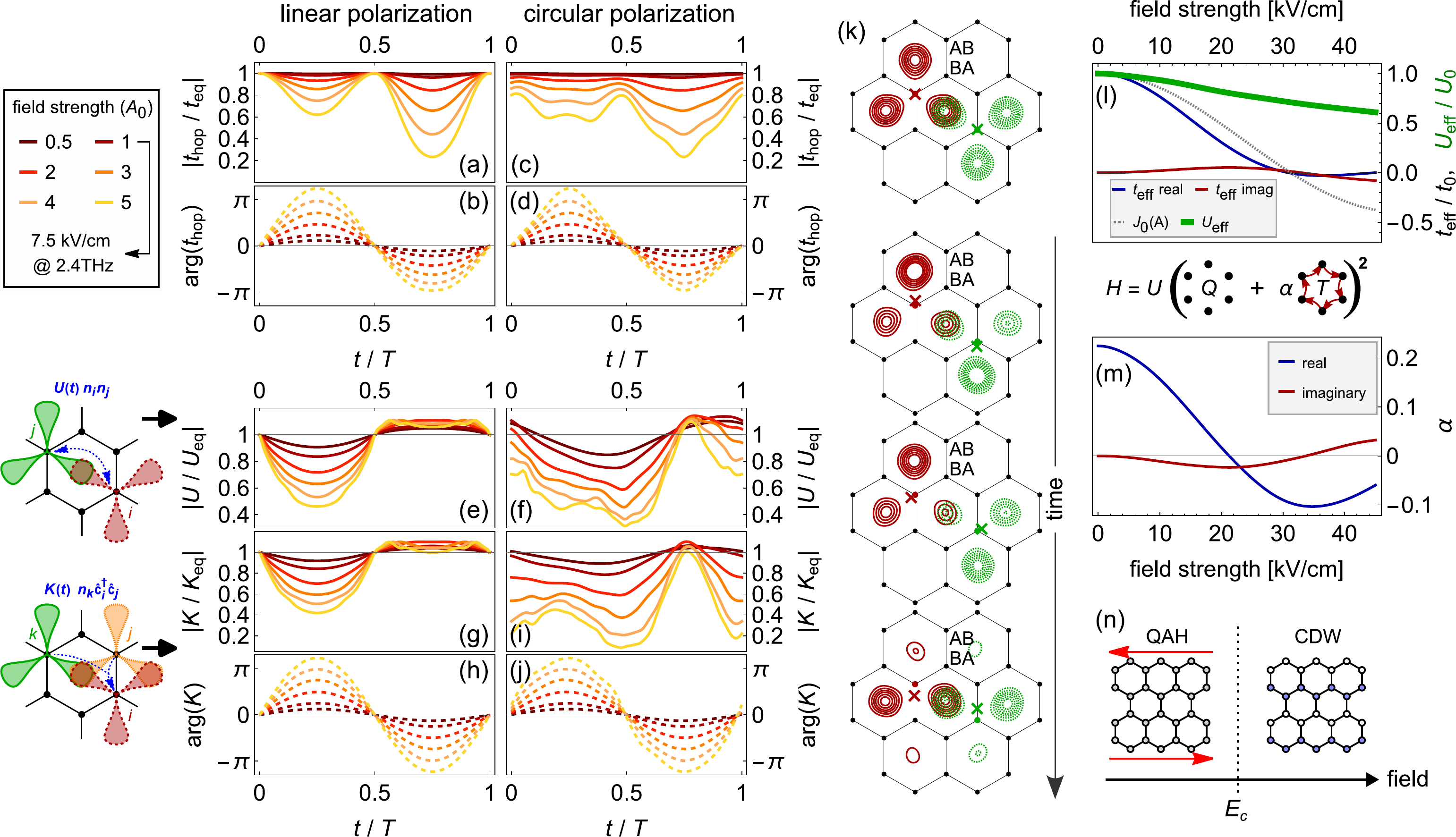}
	\caption{\captiontitle{Controlling Twisted Bilayer Graphene with Light}. (a), (b) and (c), (d) depict the light-induced time-dependent modification of nearest-neighbor hopping amplitudes $t_{\rm hop}$ due to photon-induced deformations of the fidget-spinner Wannier orbitals in twisted bilayer graphene for linearly and circularly polarized light, respectively, as a function of electric field strength ($A_0 = 1$ corresponds to 7.5kV/cm at 2.4THz). While $\arg(t_{\rm hop})$ primarily reflects the Peierls phase, the hopping magnitude becomes drastically suppressed due to dynamically reduced overlap between neighboring fidget spinner orbitals. (e)-(j) depicts the corresponding photon-dressed Coulomb interactions for a single valley, exemplified for next-nearest-neighbor density interactions $U(t)$ and correlated hopping $K(t)$. Remarkably, both reveal a dynamical suppression of up to $50\%$ of their equilibrium value, which can be understood from (k) the time-dependent deformation and reduced overlap of two fidget spinner Wannier orbitals centered in AB and BA regions of the moir\'e unit cells. For broad monochromatic pulses, (l) depicts corresponding Floquet predictions for the period-averaged effective hopping and interaction matrix elements that govern the transient prethermal dynamics of driven TBG. The renormalization of $U$ as well as the relative contribution $\alpha$ of correlated hopping interactions (m) in TBG reveals a new quantum-geometric handle, by which light can directly manipulate Coulomb interactions in moir\'e materials with poorly-localized Wannier orbitals and steer correlated phases, for instance (n) across a transition at 3/4 filling between quantum anomalous Hall and a proximal charge ordered phases.}
    \label{fig:TBGparameters}
\end{figure*}

Moir\'e heterostructures of two-dimensional van der Waals materials provide a natural platform for exploring the quantum-geometric coupling between light and Wannier functions. These structures combine twist-tunable electronic bands that become almost dispersionless at small angles with strong electronic interactions and poorly localized Wannier functions, whose light-induced deformation can be expected to dominate optical driving and responses. Magic angle twisted bilayer graphene constitutes a particularly illustrative example, with isolated bands with meV bandwidth at charge neutrality which form from charge pockets at the AA stacking sites [Fig. \ref{fig:TBG}(a)]. A fragile topological obstruction prevents the construction of localized Wannier functions if all emergent spatial symmetries are taken into account \cite{po2018,po2019}, however can be circumvented by implementing some symmetries in a nonlocal manner. The resulting maximally-localized Wannier functions generally have a ``fidget-spinner’’ shape \cite{koshino2018,kang2018,kang2019}; they are centered on the AB and BA stacking regions of the moire unit cell and form a honeycomb lattice, however exhibit three lobes that localize on neighboring AA charge pockets at the center of its hexagons. Starting from the Bistritzer-MacDonald continuum model for TBG \cite{bistritzer2011}, we construct Wannier functions in equilibrium using the gauge-fixing procedure outlined in Ref. \cite{koshino2018} for the two Bloch states per valley $\tilde{\mathbf{u}}_{n\k}(\r)$ [see Appendix \ref{appendix:tbg}]. The pair of AB/BA centered equilibrium Wannier functions is depicted in Fig. \ref{fig:TBG}(b) for a single valley. As a consequence of the three-lobe orbital structure, and even for well-screened Coulomb interactions, the resulting effective Hamiltonian
\begin{align}
    \Ham = -t_{\rm eq} \sum_{\left<ij\right>l\sigma} \CD{il\sigma} \C{jl\sigma} + U \sum_{\varhexagon} \left( \hat{Q}_{\varhexagon} + \alpha \hat{T}_{\varhexagon} - 4 \right)^2  \label{eq:KangVafek}
\end{align}
must minimally include an effective Coulomb repulsion $U$ between \textit{all} hexagon-adjacent AB / BA centered orbitals, described via a hexagon charging contribution $\hat{Q}_{\varhexagon} = \sum_{jl\sigma} \ND{j,l\sigma} / 3$ as well as correlated two-particle hopping around hexagons $\hat{T}_{\varhexagon} = i \sum_{jl\sigma} (-1)^l \CD{j+1,l\sigma} \C{j,l\sigma} + \hc$ where $l$ is the layer index \cite{kang2019,daliao21,chen21}. 

Coupling to light now deforms both the shape and center of the AB and BA fidget-spinner Wannier orbitals, which can be computed from Eq. (\ref{eq:timeDependentWannierMultiband}) and are depicted for a single valley in Fig. \ref{fig:TBG}(c) as a function of time for a single pump period, for both linear and circular polarization. This deformation depends solely on the quantum geometry of the TBG bands via the non-Abelian intravalley Berry connection, determined by the gauge-fixed Bloch states that yield the original Wannier function in equilibrium. We choose a readily attainable field strength of 15 kV/cm at 2.4THz; the pump frequency does not affect the Wannier orbital dynamics but must be chosen not to place higher-energy bands into resonance with the bands near charge neutrality. Remarkably, the weak field already yields a pronounced shift of the Wannier center, in equal and opposite directions for the AB and BA site Wannier functions. The direction is set by the sign of the non-Abelian Berry connection, projected in the direction of the gauge field. Conversely, for circular polarization, the Wannier centers precess about the respective honeycomb lattice sites with a handedness that reflects the chirality of the underlying Bloch state.

The deformation of the TBG fidget-spinner orbitals immediately entails drastic modifications of the effective electronic hopping and Coulomb interactions. First, consider the time-dependent magnitude and phase of hopping amplitudes, depicted in Fig. \ref{fig:TBGparameters}(a) and (b) for a vertical nearest-neighbor bond between AB and BA sites, for linear polarization along the vertical axis. While the phase predominantly reflects the Peierls phase $e^{i \Agauge(t) \cdot (\R_{\rm AB} - \R_{\rm BA})}$, surprisingly the magnitude of the hopping matrix element becomes rapidly quenched with increasing field strength due to the out-of-phase oscillation of the Wannier centers of the AB and BA time-dependent Wannier orbitals, which reduces their overlap. A similar situation arises for circular polarization, depicted in Fig. \ref{fig:TBGparameters} (c) and (d).

Crucially, the effective Coulomb interactions that govern the correlated phase now become dressed by light as well and oscillate with the field. This contribution dominates the coupling of magic-angle TBG to light for strong interactions and constitutes the key observation of this work. Fig. \ref{fig:TBGparameters}(e) and (f) illustrates this effect by example of the density interaction between two AB and BA sites, mediated via overlapping orbital lobes at the AA region in the shared hexagon's center. Remarkably, the interaction is coherently suppressed by $\sim 50\%$ already for comparatively weak THz fields of 40kV/cm, with similar observations for circular polarized light [Fig. \ref{fig:TBGparameters}(f)]. The root cause can be readily inferred from time snapshots of the overlaps of the two AB, BA field-dressed Wannier functions shown in Fig. \ref{fig:TBGparameters}(k) for circular polarization, which show that the weight of both AB and BA centered Wannier functions is shifted away from the shared AA region to other lobes, as a function of time.

The dynamical light-induced renormalization of correlated hopping interactions is even more pronounced, with the tie-dependent amplitude and phase depicted in Fig. [Fig. \ref{fig:TBGparameters}(g) - (j)] for linear and circular polarization. Here, the modification of the overlap of the constituent Wannier function conspires with a Peierls phase accrued by the correlated hopping process. Consequently, the pump period averaged effective correlated hopping interaction rapidly becomes suppressed already for weak fields, granting a light-induced handle to change the range and nature of effective Coulomb interactions.

The corresponding Floquet predictions for period-averaged hoppings and interactions that govern the transient steady state for wide pump pulses are depicted in Fig. \ref{fig:TBGparameters}(l) and (m) for circular polarization and a single valley, as a function of the field strength. The effective nearest-neighbor hopping is rapidly quenched and vanishes at a critical field strength to realize an ideal flat band, before reverting again to a finite bandwidth. The critical field strength $\mathcal{E}_0$ remains dictated primarily via the Peierls phase and hence closely follows expectations for Floquet dynamical localization $t_{\rm eff} \sim t_{\rm eq} J_0(A_0)$ where $J_0$ is the zeroth Bessel function and $A_0 = a_0 e \mathcal{E}_0 / \hbar \Omega$ is the dimensionless field strength with $a_0$ the moir\'e length. Importantly however, the Floquet modification of the Coulomb interactions in Eq. (\ref{eq:KangVafek}) that arises from the period-averaged deformation of the Wannier functions simultaneously yields a reduced $U_{\rm eff}$, reaching $50\%$ of its equilibrium value for a field strength of $\sim 50 kV/cm$ [Fig. \ref{fig:TBGparameters}(l)]. This intriguingly suggests that off-resonant THz radiation can be used to coherently tune magic-angle TBG across regimes $t_{\rm eff} / U_{\rm eff}$ with strong and weak correlations.

Further insight into possible light-induced phase transitions can be gleaned from inspecting the concurrent modification of the competition between density and correlated hopping interactions, which is depicted via the dimensionless parameter $\alpha$ [Eq. \ref{eq:KangVafek}] in Fig. \ref{fig:TBGparameters}(m). Starting from its equilibrium value $\sim 0.2$ \cite{kang2019,koshino2018}, $\alpha$ can be quenched to zero and flips its sign beyond a critical field strength that is distinct from kinetic dynamical localization. An immediate ramification is the possibility to dynamically access to competing phases in the rich phase diagrams that have recently been proposed for TBG to emerge from the competition between kinetics, interactions and correlated two-electron hopping \cite{kang2019,daliao21,chen21}. An intriguing example is the quantum anomalous Hall state at 3/4 filling, which competes with ferromagnetic charge order and can be stabilized via correlated hopping beyond a critical $\alpha_c \sim 0.12$. Starting from an equilibrium value $\alpha \sim 0.2$, this transition can be reached via irradiation already for field strengths $\sim 15 kV/cm$, schematically depicted in Fig. \ref{fig:TBGparameters}(n).

\sectiontitle{Discussion and Outlook}

This work introduced a new quantum geometric handle to probe and steer strongly correlated quantum materials with light, whereby light can directly deform the material's Wannier orbitals that comprise the states near the Fermi energy. This effect has profound implications for strongly interacting electron systems, leading to a concurrent light-induced modulation of electronic motion and electronic interactions that can dominate optical driving and responses in materials with non-trivial quantum geometry. Consequences for optical spectroscopy were first illustrated for a strongly interacting Hubbard chain, which can be driven across a topological phase transition to reveal a new class of domain-wall excitations whose motion and coupling to light yields an optical probe of the material's quantum geometry. We then presented ramifications for non-equilibrium control of moir\'e heterostructures and demonstrated that subjecting magic-angle twisted bilayer graphene to weak THz radiation can conspire with a fragile topological obstruction to profoundly alter the material's competing interactions. This mechanism is distinct from conventional dynamical localization, and permits tuning TBG across boundaries to competing phases.

The results of our work have important implications for theories of photon-based spectroscopies, which will need to be reevaluated in the context of correlated and topological materials \cite{Zong2023, Goulielmakis2022}. While conventional theories of probes such as second harmonic generation, Raman spectroscopy and shift currents assume that photons couple to the motion of individual electrons \cite{Devereaux2007}, the direct quantum-geometrical dressing of a Mott insulator's strong Coulomb interactions with light due to Wannier orbital deformations can readily dominate such spectroscopic responses. This can permit new insight into excitations of unconventional phases. For instance, symmetry-allowed higher-order non-local responses in Mott quantum magnets can become enhanced via photon-induced shifts of the constituent Wannier orbitals that host the local moments; an intriguing example is the polarization-resolved Raman response mediated via longer-ranged exchange processes which can probe subdominant chiral excitations in frustrated quantum magnets \cite{Shastry1990}. At the same time, the quantum-geometric coupling of light and Wannier orbitals can in principle enable new functional behavior. For instance, building on recent progress in the microscopic understanding of the bulk photovoltaic effect and shift currents in band insulators \cite{Dai2022}, an intriguing question concerns whether Mott insulators with non-trivial quantum geometry can yield new quantum-geometric design principles for obtaining enhancements to bulk photovoltaic responses mediated via light-dressed Coulomb interactions. Here, the requisite noncentrosymmetry for shift currents guarantees a non-vanishing Berry curvature for the Mott insulator’s constituent bands near the Fermi energy.

More broadly, the dynamics of topological correlated materials driven far from equilibrium via strong-field Wannier orbital modulations can yield a new route towards strong-field dressing and Floquet control \cite{Oka2019,rudner2020}. The tradeoff between realizing substantial prethermal Floquet modifications and suppressing heating to retain short-time quantum coherence remains a key challenge \cite{delatorre21}, hence an interesting question concerns whether the presented direct quantum geometric dressing of materials with strong Coulomb interactions can establish new and targeted mechanisms for optical control already for weak pump fields. Similarly, strong coupling to quantized photon modes in optical and THz cavities has recently emerged as a promising alternative to modify emergent properties of quantum materials \cite{kockum19,Schlawin2022}. It will hence be especially interesting to generalize the presented quantum-geometric coupling of light and correlated electrons to a theory of polaritonic Wannier orbitals.

Methodologically, a key application of the introduced time-dependent Wannier basis is as a faithful effective description of the coupling of strongly-interacting electrons near the Fermi energy to strong light fields, remaining independent of electromagnetic gauge choices and exact up to corrections in the inverse detuning to off-resonant higher-energy bands transitions. As the formalism relies solely on knowledge of the material’s non-Abelian Berry connections in the equilibrium Wannier basis, a promising direction concerns extracting this quantum-geometric data directly from \textit{ab initio} density functional theory (DFT) calculations, as a route towards the first principles construction of effective tight-binding descriptions of light-matter interactions for correlated electrons. An important open question concerns the role of light-induced dynamical screening corrections to the effective Coulomb interactions and their relative importance compared to the quantum-geometric contribution. These could be addressable via a time-dependent generalization of constrained random-phase approximation calculations with double counting corrections treated with appropriate care, or alternatively via computing the time-dependent Wannier functions directly from time-dependent DFT simulations of the irradiated material \cite{Runge1984, Marques2006}. The latter approach will necessitate an \textit{ab initio} implementation of the gauge fixing procedure for Wannier orbitals analogous to the one presented here, to again guarantee a faithful description of dynamics for low-frequency optical excitation. Finally, it will be interesting to generalize our quantum-geometric formalism beyond the dipole approximation to assess the role of finite-$q$ corrections.


%

\clearpage
\appendix

\section{Light-Matter Coupling in the Quantum-Geometric Gauge}\label{appendix:lightmattercoupling}

The main text presents a new quantum-geometric Bloch-Wannier basis for interacting electrons dressed by light [Eqns. (\ref{eq:geometricFieldSingleBand}), (\ref{eq:nonAbelianGeometricPhase})] that encodes a photon-induced deformation of the Wannier functions that comprise the bands near the Fermi energy which dominates optical driving and responses in strongly-correlated materials. This deformation cancels $\Delta_{n,n'}$-divergent contributions to the interband current operators between band $n$ and inert bands $n'$ that are far from resonance with the light field. To arrive at the effective light-matter-coupled low energy Hamiltonian, we start without loss of generality from the bare Hamiltonian in velocity gauge
\begin{align}
	\Ham = \Ham_0 + \Ham_{\textrm{int}}
\end{align}
with
\begin{align}
	\Ham_0 &= \int d\r ~\hat{\boldsymbol{\Psi}}^\dag(\r) \left[ \frac{1}{2m} \left[\hat{\mathbf{p}} + \Agauge(t) \right]^2 + U(\r) \right] \hat{\boldsymbol{\Psi}}(\r)  \label{eq:appendixBareHamiltonian0} \\
	\Ham_{\textrm{int}} &= \frac{1}{2} \sum_{\sigma\sigma'} \iint d\r d\r' V(\r-\r') \hat{\Psi}_{\sigma}^\dag(\r) \hat{\Psi}_{\sigma'}^\dag(\r') \hat{\Psi}_{\sigma'}(\r') \hat{\Psi}_{\sigma}(\r)   \label{eq:appendixBareHamiltonianInt}
\end{align}
with crystal potential $U(\r)$ and Coulomb interactions $V(\r-\r')$. We use a sign convention for intra- and interband Berry connections
\begin{align}
	\Aberry_{nn'}(\k) = -i \int_{\textrm{cell}} \mathbf{u}^\dag_{n,\k}(\r) \boldsymbol{\nabla}_{\k} \mathbf{u}_{n',\k}(\r)
\end{align}

\subsection{Single band}

\newcommand{\geophase}[2]{ e^{-i \hspace{-0.35cm} \int\limits_{#1}^{~~#1+\Agauge(t)} \hspace{-0.35cm} d\Aberry_{#2}} }
\newcommand{\geophaseC}[2]{ e^{i \hspace{-0.35cm} \int\limits_{#1}^{~~#1+\Agauge(t)} \hspace{-0.35cm} d\Aberry_{#2}} }

The case of a single isolated band is particularly straightforward and involves the geometric gauge ansatz described in the main text
\begin{align}
	\hat{\boldsymbol{\Psi}}(\r,t) = \sum_{n\k} e^{i\k\r} \mathbf{u}_{n,\k+\mathbf{A}(t)}(\r) ~ \geophase{\k}{nn} ~ \C{n\k}  \label{eq:appendixSingleBandField}
\end{align}
where $\Aberry_{nn}(\k)$ denotes the \textit{intraband} Berry connection for band $n$. Substituting into Eq. (\ref{eq:appendixBareHamiltonian0}), and accounting for the time-dependence of the fields $\hat{\boldsymbol{\Psi}}^\dag(\r,t) i\frac{\partial}{\partial t} \hat{\boldsymbol{\Psi}}(\r,t)$, one obtains
\begin{widetext}
\begin{align}
	\Ham_0 &= \sum_{\substack{nn' \\ \k\k'}} \int d\r~ e^{-i(\k-\k')\r} \geophaseC{\k}{nn} \mathbf{u}^\dag_{n,\k+\mathbf{A}}(\r) \left[ \frac{1}{2m} \left(\hat{\mathbf{p}} + \k' + \Agauge \right)^2 + U(\r) - i \partial_t \right] \mathbf{u}_{n',\k'+\mathbf{A}}(\r) \geophase{\k'}{n'n'} ~\CD{n\k} \C{n'\k'} \notag\\
		&= \sum_{n \k} \E_{n}(\k+\Agauge(t)) ~\CD{n\k} \C{n\k} ~-~ \sum_{\substack{nn' \\ \k\k'}}\int d\r~ e^{-i(\k-\k')\r}  \geophaseC{\k}{nn} \mathbf{u}^\dag_{n,\k+\mathbf{A}(t)}(\r) ~i\partial_t ~ \mathbf{u}_{n',\k'+\mathbf{A}(t)}(\r) \geophase{\k'}{n'n'} ~\CD{n\k} \C{n'\k'} \notag\\
		&= \sum_{n \k} \E_{n}(\k+\Agauge(t)) ~\CD{n\k} \C{n\k} ~-~ \sum_{nn'\k}  \frac{\partial \Agauge(t)}{\partial t} \cdot \left\{ \int_{\textrm{cell}} \mathbf{u}^\dag_{n,\k+\mathbf{A}(t)}(\r) ~\left[ \vphantom{\frac{}{2}} i\boldsymbol{\nabla}_{\k} + \Aberry_{n'n'}(\k + \Agauge(t)) \right] \mathbf{u}_{n',\k+\mathbf{A}(t)}(\r) \right\} \CD{n\k} \C{n'\k} \notag \\
		&= \sum_{n \k} \E_{n}(\k+\Agauge(t)) ~\CD{n\k} \C{n\k} ~+~ \sum_{nn'\k}  \frac{\partial \Agauge(t)}{\partial t} \cdot \left\{ \Aberry_{nn'}(\k+\Agauge(t)) - \delta_{n,n'} \Aberry_{nn}(\k + \Agauge(t)) \right\} \CD{n\k} \C{n'\k} \notag \\
		&= \sum_{n \k} \E_{n}(\k+\Agauge(t)) ~\CD{n\k} \C{n\k} ~-~ \sum_{n \neq n'} \sum_{\k}  \mathbf{E}(t) \cdot \Aberry_{nn'}(\k+\Agauge(t))~ \CD{n\k} \C{n'\k}
\end{align}
\end{widetext}
The first term of the transformed non-interacting Hamiltonian describes the band dispersion with the usual Peierls substitution. The second term describes light-induced interband transitions which crucially remain independent of the energy difference between bands and can therefore be neglected for detuned light fields, as described in the main text. The interacting Hamiltonian transforms as
\begin{align}
	\Ham_{\textrm{int}} &= \frac{1}{L} \sum_{\k\k'\q} V_{\k\k'\q}^{nn'm'm}(\Agauge(t))~ \CD{n,\k+\q} \CD{n',\k'-\q} \C{m',\k'} \C{m,\k}
\end{align}
with a field-dressed time-dependent interband interaction vertex
\begin{align}
	V_{\k\k'\q}^{nn'm'm} &= V(\q) \braket{u_{n,\k+\q+\Agauge(t)}}{u_{n',\k+\Agauge(t)}} \notag\\
		&\times \braket{u_{m',\k'-\q+\Agauge(t)}}{u_{m,\k'+\Agauge(t)}} \notag\\ 
		&\times~ e^{i \left[ \phi_{n}(\k+\q,t) + \phi_{n'}(\k'-\q,t) - \phi_{m'}(\k',t) - \phi_{m}(\k,t) \right]  }
\end{align}
where
\begin{align}
	 \phi_n(\k,t) = \hspace{-0.2cm}\int\limits_{\k}^{\k+\mathbf{A}(t)} \hspace{-0.3cm} \boldsymbol{\mathcal{A}}_{n}(\k') \cdot d\k'
\end{align}
is the geometric phase that describes the deformation of the corresponding Wannier functions. If the partially-filled band is energetically isolated and the light field is detuned from interband transitions, the expression quoted in the main text follows straightforwardly via discarding all interband electron-photon and Coulomb scattering terms.

\subsection{Multiband Models}

\newcommand{\Qgeo}{\hat{\mathbf{Q}}}

Consider now a set of ``entangled'' bands near the Fermi energy. The geometric gauge formalism straightforwardly generalizes by separating this set $\mathcal{G}$ from other (inert) bands $\mathcal{I}$
\begin{align}
	\hat{\boldsymbol{\Psi}}(\r,t) = \hat{\boldsymbol{\Psi}}_{\mathcal{G}}(\r,t) + \hat{\boldsymbol{\Psi}}_{\mathcal{I}}(\r,t)
\end{align}
where the entangled bands
\begin{align}
	\hat{\boldsymbol{\Psi}}_{\mathcal{G}}(\r,t) &= \sum_{n,n' \in \mathcal{G}} \sum_{\k} e^{i\k\r} \mathbf{u}_{n,\k+\mathbf{A}(t)}(\r) ~Q^{\mathcal{G}}_{nn'}(\k,t) ~ \C{n'\k}
\end{align}
transform via a non-Abelian quantum geometric phase $\Qgeo(\k,t)$, and the inert bands again transform individually, following Eq. (\ref{eq:appendixSingleBandField})
\begin{align}
		\hat{\boldsymbol{\Psi}}_{\mathcal{I}}(\r,t) &=  \sum_{n \notin \mathcal{G}} \sum_{\k} e^{i\k\r} \mathbf{u}_{n,\k+\mathbf{A}(t)}(\r) ~\geophase{\k}{nn} ~ \D{n\k}
\end{align}
The \textit{non-Abelian} quantum geometric phase for the bands of interest $\mathcal{G}$
\begin{align}
	\Qgeo(\k,t) = \hat{\mathcal{P}}~ \exp\left\{ -i \int\limits_{\k}^{\k+\Agauge(t)} d\hat{\Aberry}^{\mathcal{G}} \right\}
\end{align}
can be expressed as a momentum space path ordered ($\hat{\mathcal{P}}$) integral over their non-Abelian Berry connection $\hat{\Aberry}^{\mathcal{G}}$ and satisfies
\begin{align}
	i \partial_t \Qgeo(\k,t) = \left( \sum_\mu \hat{\Aberry}_\mu^{\mathcal{G}}(\k + \Agauge) ~\partial_t \Agauge_\mu \right) \cdot \Qgeo(\k,t) 
\end{align}
Substituting again into Eq. (\ref{eq:appendixBareHamiltonian0}), one obtains
\begin{widetext}
\begin{align}
	\Ham_0 &= \sum_{\substack{nn'mm' \in \mathcal{G} \\ \k\k'}} \int d\r~ e^{-i(\k-\k')\r} Q^\dag_{nn'}(\k,t) \mathbf{u}^\dag_{n,\k+\mathbf{A}}(\r) \left[ \frac{1}{2m} \left(\hat{\mathbf{p}} + \k' + \Agauge \right)^2 + U(\r) - i \partial_t \right] \mathbf{u}_{m,\k'+\mathbf{A}}(\r) Q_{mm'}(\k',t) ~\CD{n'\k} \C{m'\k'} \notag\\
\
	&= \sum_{\substack{nm \in \mathcal{G} \\ \k}} h_{nm}(\k, t) ~\CD{n\k} \C{m\k} - i \sum_{\substack{nn'\in \mathcal{G} \\ mm' \in \mathcal{G} \\ \k}} ~\int\limits_{\textrm{cell}} d\r~  Q^\dag_{nn'}(\k,t) \mathbf{u}^\dag_{n,\k+\mathbf{A}}(\r) ~\frac{\partial}{\partial t} \mathbf{u}_{m,\k+\mathbf{A}}(\r) Q_{mm'}(\k,t) ~\CD{n'\k} \C{m'\k'}
\end{align}
Here, the first term $h_{nm}(\k,t)$ describes the Peierls-substituted multi-band Bloch Hamiltonian, transformed into the basis of photon-dressed time-dependent Wannier orbitals
\begin{align}
	\mathbf{h}(\k,t) = \Qgeo^\dag(\k,t) ~\mathbf{h}^0(\k + \Agauge(t))~ \Qgeo(\k,t)
\end{align}
where
\begin{align}
	h_{nm}^0(\k) = \int\limits_{\textrm{cell}} \mathbf{u}^\dag_{n,\k}(\r) \left[ \frac{\left(\hat{\mathbf{p}} + \k\right)^2}{2m} + U(\r) - i \partial_t \right] \mathbf{u}_{m,\k}(\r)
\end{align}
is the equilibrium Bloch Hamiltonian. Conversely, the second term vanishes:
\begin{align}
	&- i \sum_{\substack{nn'\in \mathcal{G} \\ mm' \in \mathcal{G} \\ \k}} ~\int\limits_{\textrm{cell}} d\r~  Q^\dag_{nn'}(\k,t) \mathbf{u}^\dag_{n,\k+\mathbf{A}}(\r) ~\frac{\partial}{\partial t} \mathbf{u}_{m,\k+\mathbf{A}}(\r) Q_{mm'}(\k,t) ~\CD{n'\k} \C{m'\k'} \notag\\
	&= t\sum_{\substack{nn'\in \mathcal{G} \\ mm' \in \mathcal{G} \\ \k,\mu}} \frac{\partial A_\mu(t)}{\partial t} ~ Q^\dag_{nn'}(\k,t) \underbrace{ \left[ -i \int\limits_{\textrm{cell}} d\r~   \mathbf{u}^\dag_{n,\k+\mathbf{A}}(\r) \frac{\partial}{\partial k_\mu}  \mathbf{u}_{m,\k+\mathbf{A}}(\r) \right] }_{=~ \Aberry^{\mu}_{nm}(\k + \Agauge(t)) } Q_{mm'}(\k,t)  ~\CD{n'\k} \C{m'\k} \notag\\
	&- \sum_{\substack{nn'\in \mathcal{G} \\ mm' \in \mathcal{G} \\ \k,\mu}} \frac{\partial A_{\mu}(t)}{\partial t} \cdot \left\{ Q^\dag_{nn'}(\k,t) \underbrace{ \left[ ~\int\limits_{\textrm{cell}} d\r~   \mathbf{u}^\dag_{n,\k+\mathbf{A}}(\r) \mathbf{u}_{m,\k+\mathbf{A}}(\r) \right] }_{ =~ \delta_{n,m} } \sum_{m'' \in \mathcal{G}} \Aberry^{\mu}_{mm''}(\k + \Agauge(t))~ Q_{m''m'}(\k,t) \right\} ~\CD{n'\k} \C{m'\k} \notag\\
	&= 0
\end{align}
\end{widetext}

\section{Diagrammatic analysis of optical responses in the quantum-geometric gauge and conventional velocity gauge} \label{appendix:diagrams}

The main text presents a new diagrammatic dictionary for efficiently calculating linear and nonlinear optical responses in the quantum-geometric velocity gauge that encodes the light-induced deformation of the system's Wannier orbitals. To connect this exposition with conventional computations of optical responses in velocity gauge, this section elaborates on a formally exact diagrammatic analysis of optical absorption $\Real \sigma(\omega)$ presented in the main text in Fig. \ref{fig:diagrams}(c), demonstrating that calculations in the quantum-geometric gauge represent the formally exact gauge-invariant response, up to neglecting $1 / (\Delta - \omega)$ corrections from interband transitions that are off-resonant with detuning $\Delta - \omega$.

Choosing the quantum-geometric time-dependent Wannier basis in velocity gauge, optical conductivity of a set $\mathcal{G}$ of interacting energetically-isolated bands is governed by four Feynman diagrams depicted in Fig. \ref{fig:diagrams}(c), in addition to imaginary diamagnetic contributions. We focus here on \textit{low-frequency} optical absorption, off resonance from interband transitions; the imaginary response readily follows via Kramers-Kronig relations. At photon frequencies $\omega \ll \Delta$ the gap to higher-energy conduction bands or deeper fully-filled valence bands, the final state after an absorption event cannot generate any interband excitations. This can be straightforwardly expressed via the Fermi golden rule expression
\begin{align}
    \Real \sigma(\omega) = \frac{2\pi}{\omega} \sum_{j \in \mathcal{G}} \left| \bra{j} \hat{J}(\omega) \ket{0}  \right|^2 \delta(\omega + E_0 - E_j)  \label{eq:fermiGoldenRuleConductivity}
\end{align}
where $j \in \mathcal{G}$ denotes the set of many-body states which account for excitations solely within the partially-filled bands near the Fermi energy, with all inert bands $\notin \mathcal{G}$ fully filled or empty.

In ordinary velocity gauge approaches for non-interacting quantum systems, the current operator $\hat{J}(\omega) \equiv \hat{J}^{\rm intra}$ involves solely the ``intraband'' current operator acting on the subset of bands $n,n' \in \mathcal{G}$ near the Fermi energy
\begin{align}
    \hat{J}^{\rm intra} = \sum_{nn'\k} j_{\k}^{nn'} \CD{n\k} \C{n'\k}
\end{align}
where electron spin is again included in band indices $n$.

Crucially however, and as described in the main text, the \textit{interband} current operator $j_\k^{n,l}$ for transitions to detuned inert bands $l \notin \mathcal{G}$ scales with the band gap if the bands have non-trivial quantum geometry $\Aberry_\k^{n,l} \neq 0$. However, placing an electron in a higher-energy inert band immediately places an energy cost $\sim \Delta$ on such photo-excited intermediate states. Hence, to leading order $\mathcal{O}(1)$ in the inverse detuning of interband transitions, the photo-excited intermediate state must immediately recombine and scatter back to the original band near the Fermi energy. In strongly-interacting quantum materials, the dominant scattering contribution comes from Coulomb scattering, and entails a higher-order contribution to the \textit{effective} current operator. Denoting intermediate many-body states with a single photo-excited electron in band $l$ as
\begin{align}
    \ket{j', l\k} \equiv \ket{j'} \otimes \DD{l\k} \ket{0_l}
\end{align}
with $\ket{j'}$ again a many-body state for excitations solely in $\mathcal{G}$ and $\DD{l\k}$ the fermion creation operator for an electron in an inert higher-energy band, these processes can be included in Eq. (\ref{eq:fermiGoldenRuleConductivity}) via
\begin{align}
    \hat{J}(\omega) &= \hat{J}^{\rm intra} + \hat{J}^{\rm inter}_{\rm eff}(\omega) + \mathcal{O}(1/\Delta)
\end{align}
where
\begin{align}
        \hat{J}^{\rm inter}_{\rm eff}(\omega) &= \sum_{j'n l\k} \frac{ \hat{V}_{\rm inter} \ketbra{j',l\k}{j',l\k} \left( \Delta_\k^{ln} \Aberry_\k^{ln} \DD{l\k} \C{n\k} \right) }{\omega + E_0 - E_{j'} - \Delta_{\k}^{ln} + i\eta} \notag\\
\
        &- \sum_{j'n l\k} \frac{ \left( \Delta_\k^{ln} \Aberry_\k^{nl} \DD{l\k} \C{n\k} \right) \ketbra{j',l\k}{j',l\k} \hat{V}_{\rm inter}^\dag }{\omega + E_0 - E_{j'} - \Delta_{\k}^{ln} + i\eta}\label{eq:interbandcurrentFGR}
\end{align}
where $\Aberry_\k^{ln}$ is the interband Berry connection and
\begin{align}
    \hat{V}_{\rm inter} = \sum_{\substack{\k\k'\q \\ n_1 n_2 n_3 l}} V_{\k\k'\q}^{n_1n_2n_3l} \CD{n_1,\k+\q} \CD{n_2,\k'-\q} \C{n_3,\k'} \D{\k,l}
\end{align}
with 
\begin{align}
    V_{\k\k'\q}^{n_1n_2n_3l} = \braket{u_{n_1,\k+\q}}{u_{l,\k}} \braket{u_{n_2,\k'-\q}}{u_{n_3,\k'}}
\end{align}
the interband Coulomb scattering matrix element that scatters one photo-excited electron in inert band $l$ off an electron in band $n_3$ near the Fermi energy, leaving both electrons in $\mathcal{G}$. 

To recover the results derived in the quantum-geometric gauge of time-dependent Wannier functions, suppose now that the band gaps $\Delta^{nl}$ to inert bands are much larger than the photon frequency. In this case, Eq. (\ref{eq:interbandcurrentFGR}) simplifies drastically
\begin{align}
    \hat{J}^{\rm inter}_{\rm eff}(\omega) &= \hspace{-0.3cm} \sum_{\substack{\k\k'\q \\ n_1n_2n_3n}} \hspace{-0.3cm} T_{\k\k'\q}^{n_1n_2n_3n} \CD{n_1,\k+\q} \CD{n_2,\k'-\q} \C{n_3,\k'} \C{n\k}
\end{align}
with
\begin{align}
    T_{\k\k'\q}^{n_1n_2n_3n} &= \braket{u_{n_2,\k'-\q}}{u_{n_3,\k'}} \sum_{l \notin \mathcal{G}} \bra{0_l} \D{l\k} \DD{l\k} \ket{0_l} \notag\\
    &\times  \left( -\braket{u_{n_1,\k+\q}}{u_{l,\k}} \Aberry_\k^{ln} ~+~ \Aberry_{\k+\q}^{n_1,l} \braket{u_{l,\k+\q}}{u_{n,\k}}  \right)
\end{align}
As $l$ formally sums over \textit{all} inert bands $\bra{0_l} \D{l\k} \DD{l\k} \ket{0_l} = 1$ that are not included in the set of partially-filled bands $\mathcal{G}$, the second line can be simplified drastically via using 
\begin{align}
    \sum_{l \notin \mathcal{G}} \ketbra{l}{l} = 1 - \sum_{n \in \mathcal{G}} \ketbra{n}{n}
\end{align}
and $\Aberry_{\k}^{nl} = -i\braket{u_{n\k}}{\partial_\k u_{l\k}} = +i \braket{\partial_\k u_{n\k}}{u_{l\k}}$, which yields
\begin{align}
    &T_{\k\k'\q}^{n_1n_2n_3n} = \braket{u_{n_2,\k'-\q}}{u_{n_3,\k'}} \notag\\
    &\hspace{1.6cm} \times (-i) \sum_{l \notin \mathcal{G}} \left\{ \braket{u_{n_1,\k+\q}}{u_{l,\k}} \braket{u_{l,\k}}{\partial_\k u_{n,\k}} \right. \notag\\
    & \hspace{1.8cm} + \left. \braket{\partial_\k u_{n_1,\k+\q}}{u_{l,\k+\q}} \braket{u_{l,\k+\q}}{u_{n,\k}} \right\} \notag\\
    =& \braket{u_{n_2,\k'-\q}}{u_{n_3,\k'}} \left\{ \vphantom{\frac{1}{2}} (-i \partial_\k) \braket{u_{n_1,\k+\q}}{u_{n,\k}} \right. \notag\\
    &+ \left. \sum_{n' \in \mathcal{G}} \left( \braket{u_{n_1,\k+\q}}{u_{n',\k}} \Aberry_\k^{n'n} - \Aberry_{\k+\q}^{n_1,n'} \braket{u_{n',\k+\q}}{u_{n,\k}} \right) \right\}
\end{align}
This recovers precisely the paramagnetic current operator for a set of bands $\mathcal{G}$ that was derived using the quantum-geometric gauge in the main text [Eq. (\ref{eq:twoparticleCurrentVertex})]. The derivation demonstrates rigorously that the quantum-geometric gauge introduced in this manuscript permits an asymptotically exact description of low-frequency optical responses and non-equilibrium driving in correlated materials, up to ignoring off-resonant interband transitions that are suppressed with the inverse detuning. For conciseness, the above derivation implicitly assumed that all inert bands $l$ are unoccupied conduction bands; the results remain unchanged upon also accounting for fully-filled deep valence bands.

\section{Quantum-Geometric Excitations in an Interacting Topological 1D Chain} \label{appendix:1Dchaintoymodel}

\subsection{Wannier Functions}

The maximally-localized Wannier functions for the lower band $\ket{\varphi_R} = \frac{1}{\sqrt{L}} \sum_k e^{i k R} \ket{u_-(k)}$ can be constructed from the Bloch state $\ket{u(k)}$ with a gauge choice
\begin{align}
	\ket{u(k)} = \frac{\left[\begin{array}{c} \sqrt{1+\lambda} \left(1 + e^{-i k} \right) \\ \sqrt{1-\lambda} \left( 1 - e^{-i k} \right) \end{array}\right]}{2 \sqrt{(1 + \lambda \cos(k)) } }  \label{eq:BlochState1DFlatBand}
\end{align}
The band dispersion $-2\epsilon_0 \cos(k)$ is independent of $\lambda$, permitting the independent tuning of the band width (nearest-neighbor hopping between Wannier orbitals) and the spatial extent of the Wannier orbitals using parameters $\epsilon_0$ and $\lambda$, respectively. The usual gauge-invariant lower bound bound for the Wannier spread $\left< x^2 \right>$ via the Fubini-Study metric $g(k)$ averaged over the Brillouin zone can be evaluated exactly and yields $\int g(k) dk/2\pi = 1 / 4 \sqrt{1 - \lambda^2}$. The extent becomes minimal for $\lambda = 0$, and diverges as the system approaches $\lambda \to \pm 1$. Fig. \ref{fig:toymodel}(c) in the main text shows the real-space localization of the Wannier function in the trivial and topological phase, for representative parameters. Notably, the Wannier function becomes perfectly localized on a bond and with weight only on the two neighboring sites if $\lambda = 0$. 

\subsection{Exact Bond Density Wave Ground State}\label{appendix:exactCDWsolution}

Interestingly, the interacting half-filled flat band (corresponding to overall quarter filling) admits an exact bond density wave (BDW) ground state. To see this, note that the flat band Hamiltonian consists of a sum of positive-semidefinite terms upon shifting the flat band to zero energy. Upon closer inspection, this permits constructing an exact zero-energy ground state that is simultaneously annihilated by electronic hopping and the Hubbard interaction. The former is immediately true for any wave function $\ket{\Psi} = \DD{k_1} \DD{k_2} \cdot \ket{0}$ where $\DD{k}$ creates an electron in the zero-energy flat band.

To find simultaneous zero-energy eigenstates of the Coulomb repulsion, we rewrite the interaction as
\begin{align}
    \Ham_{\rm int} &= U \sum_{q} \hat{P}^\dag_q \hat{P}_q
\end{align}
where
\begin{align}
    \hat{P}_q &= \frac{1}{2\sqrt{L}} \sum_k \left( \C{-k+q,p} \C{k,s} - \C{-k+q,s} \C{k,p} \right)
\end{align}
annihilates a pair of electrons in an $s$ and a $p$ orbital. Crucially, $\hat{P}^\dag_q \hat{P}_q$ is a positive semidefinite operator. Therefore, a zero-energy ground state must be simultaneously annihilated by \textit{each} $\hat{P}_q$ for all $q$. Since we already require the ground state to carry only electrons in the flat band, we can expand $\C{k,s} = u_{k,s} \D{k} + u_{k,p} \OP{f}{k}$ with Bloch states $\mathbf{u}_k$ defined in Eq. (\ref{eq:BlochState1DFlatBand}), and drop the annihilation operator acting on the conduction band. We arrive at a set of conditions
\begin{align}
    \sum_{k} \left( u_{-k+q,p} u_{k,s} - u_{-k+q,s} u_{k,p} \right) \D{-k+q} \D{k} \ket{\Psi} = 0 ~~ \forall~ q   \label{eq:ToyModelPairAnnihilationConstraints}
\end{align}

At half filling of the flat band, a zero-energy ground state that satisfies these constraints can be readily constructed and reads
\begin{align}
    \ket{\Psi_{\textrm{BDW}}} =  \prod_{0\leq k < \pi}\frac{1}{\mathcal{N}_k} \left( \alpha_{k}~ \DD{k} + \alpha_{k+\pi}~ \DD{k+\pi} \right) \ket{0}  \label{eq:ToyModelGroundStateAnsatz}
\end{align}
where $\mathcal{N}_k = \sqrt{|\alpha_k|^2+|\alpha_{k+\pi}|^2}$ is a normalization factor, and
\begin{align}
    \alpha_k = \sqrt{1 + \lambda\cos(k)}
\end{align}
This exact many-body ground state describes a bond density wave with spontaneously broken translation symmetry and ordering wave vector $q = \pi$. This ground state is unique, up to its partner symmetry-breaking BDW state that is shifted by a lattice translation.

Further insight can be gained for $\lambda = 0$, for which the ground state becomes especially simple
\begin{align}
    \ket{\Psi_{\textrm{BDW}}(\lambda=0)} =  \prod_{R~ \in~ \substack{\textrm{even} \\ \textrm{sites}}} \DD{R} \ket{0}  \label{eq:ToyModelGroundStateLambda1}
\end{align}
and forms a semiclassical bond ordering pattern, where $\DD{R}$ is the creation operator for a bond-centered Wannier orbital.

\subsection{Domain Wall Excitations} \label{appendix:CDWinteractionperturbation}

Suppose that the lower band is half filled while the upper band is empty, and interactions $U \ll M$ are much weaker than the energy gap to the conduction band. In this case, the low-energy physics should be dictated solely via electronic dynamics in the valence band. These can be determined straightforwardly via projection of the interacting Hamiltonian into a basis of valence-band Wannier orbitals.

Starting from a local Hubbard interaction between orbitals, interaction matrix elements in the Wannier basis can now be computed as
\begin{align}
    V_{R_1R_2R_3R_4} = \frac{U}{2} \sum_{R} \braket{\varphi_{R-R_1}}{\varphi_{R-R_4}} \braket{\varphi_{R-R_2}}{\varphi_{R-R_3}}  \label{eq:appendixWannierInteraction1D}
\end{align}

For small $\lambda$, the Wannier function decays rapidly with distance $R$. The short-ranged matrix elements can be evaluated exactly $\braket{\varphi_{R=0}}{\varphi_{R=0}} = \frac{4 e_0 \left(\lambda ^2-1\right) k_0-2 e_0^2 (\lambda -1)+2 (\lambda +1) k_0^2}{\pi ^2 \lambda ^2} \approx \frac{1}{2} + \mathcal{O}(\lambda^2)$ and $\braket{\varphi_{R=0}}{\varphi_{R=1}} = \frac{2 \left(e_0^2 (\lambda -1)+(\lambda +1) k_0^2\right)}{\pi ^2 \lambda } \approx -\frac{\lambda}{4} + \mathcal{O}(\lambda^3)$, where $k_0 = K(\frac{2\lambda}{\lambda - 1})$ and $E_0 = E(\frac{2\lambda}{\lambda - 1})$ with $K$, $E$ the elliptic integrals of the first and second kind. Longer-ranged matrix elements scale as a higher power of $\lambda$; the interacting Hamiltonian in Wannier basis presented  in Eq. (\ref{eq:perturbedHamiltonianU}) in the main text now follows from expanding the sum over $R$ in Eq. (\ref{eq:appendixWannierInteraction1D}) using the matrix elements above, to linear order in $\lambda$.

\section{Numerical Simulation} \label{appendix:simulation}

The optical conductivity comparisons in Fig. \ref{fig:toymodel}(f) in the main text between the full two-band model of atomic orbitals and the single-band Wannier model are performed using large-scale exact diagonalization for a chain length $L = 14$ and Wannier spread $\lambda = -0.1$, with spectroscopic response functions computed via Lanczos iteration. The high-frequency response in Fig. \ref{fig:toymodel}(f) is computed for free electrons; we have verified that an interacting calculation yields comparable behavior up to finite-size corrections. The calculation in Fig. \ref{fig:toymodel}(j) for the single-band Wannier model is performed for chain lengths $L=26$. We change the band gap by scaling $\Delta$ while keeping $U$ constant. The high-frequency response is calculated using exact Lehmann representation on the non-interacting model. The low-frequency response plot is calculated using exact diagonalization, which offers better low-frequency resolution.

The time-evolution in Fig. \ref{fig:toymodel}(h-i) in the main text is done for a chain length $L=8$ with a sinusoidal pulse $A(t) = A_0 \sin(\omega t)$, with $A_0 = 2.5U$ and $\omega = U/4$. To compare the validity of the theory of light-deformed Wannier functions to usual velocity gauge calculations in a truncated band basis, we furthermore simulate the latter using the projection of the full time-dependent two-band Hamiltonian into the lower flat band, which yields
\begin{align}
    \Ham_{\text{naive}}(t) =& \sum_k u^*_{k\alpha} h_{\alpha\beta} (k+A(t)) u_{k\beta} \DD{k} \D{k} \notag \\
    &+ \frac{U}{L} \sum_{\substack{kpq \\ \alpha \beta =\pm 1}} u^*_{k+p\alpha} u^*_{p-q\beta} u_{p\beta} u_{k\alpha}\DD{k+q} \DD{p-q} \D{p} \D{k} 
\end{align}
and reveals drastic discrepancies as discussed in the main text, which vanish only in the limit of perfectly-localized Wannier functions in the trivial atomic limit.

\section{Light-Dressed Wannier Functions in Twisted Bilayer Graphene}\label{appendix:tbg}

The Wannier functions for twisted bilayer graphene in the main text are computed starting an effective continuum theory for Dirac fermions. Neglecting intervalley scattering, the Hamiltonian per valley can be written as \cite{bistritzer2011,koshino2018}
\begin{align}
    H_\nu(\k) = \left[\begin{array}{ll} H_{1,\nu} & U_\nu^\dag \\ U_\nu & H_{2,\nu} \end{array}\right]
\end{align}
where $H_{l,\nu} = -\hbar v_D \hat{R}(\theta) \left(\k - \mathbf{K}^l_\nu\right) \cdot [\nu \hat{\sigma}_x,~ \hat{\sigma}_y]$ describes the Dirac fermion with velocity $v_D$ for valley $\nu$ in layer $l$ with rotation matrix $\hat{R}$ and Pauli matrices $\hat{\sigma}$, and
\begin{align}
    U_\nu &= \left[\begin{array}{cc} w_0 & w_1 \\ w_1 & w_0 \end{array}\right] + e^{i\nu \mathbf{b}_1 \cdot \mathbf{r}} \left[\begin{array}{cc} w_0 & w_1 e^{-2\pi \nu i/3} \\ w_1 e^{2\pi \nu i/3} & w_0 \end{array}\right] \notag\\
    &+ e^{i\nu (\mathbf{b}_1 + \mathbf{b}_2) \cdot \mathbf{r}} \left[\begin{array}{cc} w_0 & w_1 e^{2\pi \nu i/3} \\ w_1 e^{-2\pi \nu i/3} & w_0 \end{array}\right]  
\end{align}
describes the effective interlayer coupling. Here, $\mathbf{K}^l_\nu$ denotes the location of the Dirac points $\nu$ in layer $l$, and $\mathbf{b}_1$, $\mathbf{b}_2$ are the moir\'e lattice vectors. Using $\hbar v_D / a = 2.1354 {\rm eV}$, $w_0 = 0.0797 {\rm eV}$ and $w_1 = 0.0975 {\rm eV}$, the Wannier functions in equilibrium are calculated starting from the gauge fixing procedure for Bloch wave functions presented in Ref. \cite{koshino2018}, followed by optimization of Wannier spread functional, which yields a pair of Bloch-Wannier functions $\tilde{\mathbf{u}}_{m\k}(\r)$ with a smooth phase choice.

As described in the main text, the light-induced time-dependent deformation of the resulting fidget-spinner Wannier functions is governed by a non-Abelian quantum-geometric phase that parallel transports Bloch states from $\k+\Agauge(t)$ to $\k$. Starting from  sampling $\tilde{\mathbf{u}}_{m\k}(\r)$ on a $144 \times 144$ momentum point grid in the moir\'e Brillouin zone, we compute the non-Abelian phase factor $\mathbf{Q}(\k+\Agauge(t),\k)$ via a finite-difference approximation of the momentum space path ordered integral by defining
\begin{align}
    \mathbf{U}(\k+\delta\k, \k) = \hat{\mathcal{P}} e^{i \hspace{-0.1cm} \int\limits_{\k}^{\k+\delta\k}  \hspace{-0.1cm} \Aberry(\k')\cdot d\k'} \approx e^{\frac{\mathbf{M}(\k+\delta\k,\k) - \mathbf{M}^\dag(\k+\delta\k,\k)}{2}}
\end{align}
where $\mathbf{M}(\k+\delta\k,\k) = \log \mathbf{D}(\k+\delta\k,\k)$ with $D_{mm'}(\k,\k') = \braket{ \tilde{\mathbf{u}}_{m,\k} }{ \tilde{\mathbf{u}}_{m,\k'} }$ the $2 \times 2$ overlap matrix of Bloch states for a momentum shift $\delta \k$ on the sampled $\k$ point grid. The non-Abelian phase can now be evaluated as $\mathbf{Q}(\k+\Agauge(t),\k) = \mathbf{U}(\k+\Agauge(t),\k+\Agauge(t)-\delta\k) \cdots \mathbf{U}(\k+2\delta\k,\k+\delta\k) \cdot \mathbf{U}(\k+\delta\k,\k)$ with the magnetic vector potential $\Agauge(t)$. As discussed in the main text, the choice of momentum space path to connect $\k$ and $\k + \Agauge(t)$ remains a residual gauge freedom of the time-dependent Wannier basis. A particularly simple choice is an L shaped path that first parallel transports the Bloch states in the $k_x$ direction with displacement $\Agauge_x$, followed by parallel transport along the $k_y$ direction with displacement $\Agauge_y$; we note that for the weak fields and Bloch-Wannier states with smooth non-Abelian Berry connections considered in this work, discrepancies between path choices are negligible. The resulting time-dependent Wannier function now readily follows via a Fourier transform
\begin{align}
    \varphi_{m,\R,i}(\r,t) = \sum_{\k m'} e^{i\k (\r-\R)} Q_{mm'}(\k-\Agauge(t), \k) \tilde{\mathbf{u}}_{m'\k}(\r)
\end{align}

\subsection{Time-Dependent Hopping Amplitudes}

The gauge-fixed Bloch-Wannier states $\tilde{\mathbf{u}}_{m\k}(\r)$ define the equilibrium Bloch Hamiltonian $h^{\rm eq}_{mm'}(\k) = \bra{\tilde{\mathbf{u}}_{m\k}} H_\nu(\k) \ket{\tilde{\mathbf{u}}_{m\k}}$ per valley $\nu$. Armed with the geometric phases $\mathbf{Q}(\k+\mathbf{A}(t),\k)$, the light-dressed Bloch Hamiltonian readily follows as
\begin{align}
    \mathbf{h}(\k,t) = \mathbf{Q}(\k,\k+\mathbf{A}(t)) ~\mathbf{h}^{\rm eq}(\k + \mathbf{A}(t))~ \mathbf{Q}(\k+\mathbf{A}(t),\k)
\end{align}
with real-space hopping matrix elements computed via Fourier transform.

\subsection{Photon-Dressed Coulomb Interactions}

Light-dressed time-dependent Coulomb interactions and hopping matrix elements described in the main text are readily evaluated using the time-dependent Wannier basis described above. Coulomb interactions
\begin{align}
    V_{\R_1\R_2\R_3\R_4}^{m_1m_2m_3m_4}(t) &= \iint d\r d\r'~ \sum_{ij} \varphi_{m_1\R_1,i}^\star(\r,t) \varphi_{m_2\R_2,j}^\star(\r',t) \notag\\
        &\times ~U(|\r-\r'|) \varphi_{m_3\R_3,j}(\r,t) \varphi_{m_4\R_4,i}(\r,t) \notag\\
        &\times e^{i \left( \R_1 + \R_2 - \R_3 - \R_4 \right) \cdot \Agauge(t)}
\end{align}
where $U(r)$ is the Coulomb interaction, and $\varphi_{m,\R,i}(\r,t)$ is the time-dependent Wannier orbital $m$ for unit cell $\R$ with sublattice index $i$, which is a function of the light field $\Agauge(t)$. We adopt a screened interaction that accounts for the proximal metallic gates, with $U(r) = \sum_{n = -\infty}^{\infty} \frac{ (-1)^n }{\sqrt{n^2 + (r/\xi)^2}}$ \cite{throckmorton12}. With gate distances typically close to the moir\'e lattice spacing, the screening length $\xi$ is chosen to be equal to the lattice constant without loss of generality; the qualitative behavior of the light-induced modulation of Coulomb interactions relative to their equilibrium values remains largely independent of $\xi$.

\end{document}